\documentclass[intlimits,twoside,a4paper]{article}

\usepackage{amsmath,amssymb}
\usepackage{graphicx}

\usepackage{color}
\usepackage[T2A]{fontenc}
\usepackage[cp1251]{inputenc}

\usepackage[eqsecnum]{cmpj2}

\issue{2015}{18}{1}{13607}
\doinumber{10.5488/CMP.18.13607}

\title[What is liquid in random porous media]%
{What is liquid in random porous media: the Barker-Henderson perturbation theory%
\thanks{Dedicated to Prof. Douglas Henderson on the occasion of his 80th birthday.}}

\author[Holovko et al.]{M.F.~Holovko, T.M.~Patsahan, V.I.~Shmotolokha}
\address{Institute for Condensed Matter Physics of the National Academy of
Sciences of Ukraine, 1~Svientsitskii~St., 79011 Lviv, Ukraine}

\date{Received March 3, 2015}

\begin{document}

\maketitle

\begin{abstract}
We apply the Barker-Henderson (BH) perturbation theory to the study of a Lennard-Jones fluid
confined in a random porous matrix formed by hard sphere particles. In order to describe
the reference system needed in this perturbation scheme, the extension of the scaled particle
theory (SPT) is used. The recent progress in the development of SPT approach for a hard
sphere fluid in a hard sphere matrix allows us to obtain very accurate results for thermodynamic
properties in such a system. Hence, we combine the BH perturbation theory with the SPT approach
to derive expressions for the chemical potential and the pressure of a confined fluid.
Using the obtained expressions, the liquid--vapour phase diagrams of a LJ fluid in HS matrix are built
from the phase equilibrium conditions.
Therefore, the effect of matrix porosity and a size of matrix particles is considered.
It is shown that a decrease of matrix porosity lowers both the critical temperature and the critical
density, while the phase diagram becomes narrower. An increase of a size of matrix particles
leads to an increase of the critical temperature.
From the comparison it is observed that the results obtained from the theory are in agreement
with computer simulations. The approach proposed in the present study can be
extended to the case of anisotropic fluid particles in HS matrices.
\pacs 61.20.Gy, 61.43.Gt
\end{abstract}

\section{Introduction}

The original understanding of the nature of the liquid state of matter is connected with Van der Waals equation of state formulated nearly 150 years ago~\cite{VanderWaals88}. The Van der Waals picture focuses on different roles of the strong short-ranged repulsive and
long-ranged attractive intermolecular interaction in forming the equilibrium properties of dense fluids. According to this picture, the harsh repulsive interactions fix the shape and size of molecules and essentially determine a high-density fluid structure. While the
contribution of repulsive interaction is entropic, the contribution of attractive interaction is mainly energetic and can be treated
as the perturbation.
The first theory of liquids based on the Van der Waals idea was proposed by Barker and Henderson~(BH) nearly fifty years ago~\cite{Barker67,BarkerHen67,Barker76}. Within this theory, the intermolecular potential is separated into the repulsive and attractive parts. The short-ranged repulsive contribution is described
within the framework of the hard spheres model while the long-ranged attractive part is included within the Zwanzig high-temperature perturbation theory~\cite{Zwanzig54}.
A few years later, Andersen, Chandler and Weeks~(ACW) developed a somewhat different theory of liquid based on the Van der Waals
approach~\cite{Weeks71,Andersen72,Andersen76,Andersen1972}. In this theory, instead of intermolecular interactions,
they separated intermolecular forces into repulsive and attractive parts. They also used the optimized cluster expansion~(OCE)
instead of the high-temperature perturbation theory for the treatment of attractive forces.
However, the first term of perturbation related to the high temperature approximation~(HTA) in both theories is identical except
that the repulsive and attractive interactions are not exactly the same.

In this paper we make use of the Van der Waals ideas and expand BH theory for the description of liquids adsorbed in random porous media.
To this end, we use the Madden-Gland model~\cite{Madden90}. According to this model, a porous medium is presented as a quenched configuration of randomly distributed hard spheres forming a so-called matrix.
A specific description of a fluid in such porous media is connected with double quenched-annealed averages: the annealed average
is taken over all fluid configurations and the additional quenched average should be taken over all realizations of a matrix.
One of the popular approaches to the solution of this  problem is based on the replica method, which allows for the extension of many theoretical
methods of liquid state physics to the case of a fluid confined in a random porous medium.
For instance, using the replica Ornstein-Zernike~(ROZ) integral equation theory~\cite{Given92}, the statistical mechanics approach of liquid state
was extended to the description of various fluids confined in random porous media~\cite{Rosinberg99,Pizio2000}, including the chemically
reacting fluids~\cite{Trokh96,Trokh97}. However, unlike bulk fluids, no analytical result has been obtained from the ROZ integral equation approach
even for such a simple model as a hard sphere~(HS) fluid in a~HS matrix. At the same time, the model of HS fluid has a peculiar importance, since similar to the case of a bulk fluid~\cite{Yukhnovski80,Hansen06}, it can be used as the reference system in the development of different perturbation schemes.

The first rather accurate analytical results for a HS fluid in a HS matrix were obtained quite recently in~\cite{Hol09,Chen10,Pat11,Hol12,Hol13,Hol14} by extending the scaled particle theory~(SPT)~\cite{Reiss59,Reiss60} to the case of a HS fluid confined in random porous media.
The SPT approach is based on the combination of the exact treatment of a point scaled particle in a HS fluid combined with the thermodynamical treatment of a macroscopic scaled particle. The exact result for a point scaled particle in a HS fluid in a random porous medium was obtained in~\cite{Hol09}. However, the approach
proposed in~\cite{Hol09} referred to as SPT1 contains a subtle inconsistency appearing when the size of matrix particles is much larger than the size of fluid particles. Later on, this inconsistency was eliminated in a new approach referred to as SPT2~\cite{Pat11}. Starting from this formalism, a series of new approximations were developed~\cite{Pat11,Hol12,Hol13,Hol14}. Among these approximations we select only SPT2b and SPT2b1, which will be used in this paper for the description of the reference system in the BH perturbation theory.

The liquid--vapour phase diagrams play a central role in understanding the nature of liquids. In contrast to the bulk case, the results of investigations of liquid--vapour phase equilibrium of a simple fluid in random porous media are rather controversial. Computer simulations of a simple fluid confined in a HS matrix~\cite{Page96,Page1996,Alvarez99,Brennan02} demonstrate a possibility of the existence of two phase transitions. One is analogous to the bulk liquid-vapor transition, but with a narrower coexistence curve and lower values of the critical density and critical temperature. The second transition occurs at lower temperatures and at higher densities, and it is interpreted as a phenomenon related to the wetting effects in the fluid located in more confined regions of the matrix.
On the other hand, in more thorough investigations~\cite{Brennan02}, it was noticed that this second transition is extremely sensitive to a particular matrix configuration. In the case of HS matrix, it was shown that two phase transitions appear in some realizations of the matrix, while a single phase transition was observed in the others. In order to explain this observation, optimized cluster expansions were used in~\cite{Kierlik97}. It was found that different approximations can lead to qualitatively different results.
For example, the mean spherical approximation~(MSA) gives only a single liquid--vapour transition, but the inclusion of the second and the third
cluster coefficients result in two phase transitions. At the same time, it was shown in~\cite{Pat03} that the association theory leads to one phase transition. Our recent investigation, in which we generalized the Van der Waals equation for simple fluids in random porous media, also demonstrates only one liquid--vapour transition~\cite{Hol13}.

In the present study we combine the BH theory with the previously developed SPT approach to describe a liquid--vapour phase behaviour
of a Lennard-Jones fluid confined in a random matrix formed by hard spheres (HS matrix). For a comparison, the results obtained in different approximations are considered as well. As it was mentioned above for the reference system, the SPT approach is applied using the SPT2b approximation and its improved version SPT2b1. It should be noted that the SPT2b approximation is our first really successful result for confined fluids~\cite{Pat11}, although it may have essential problems at high fluid densities and/or for low matrix porosities. At the same time, the improved SPT2b1 approximation is based on the original SPT2b, but it is free of this shortcoming and gives an accurate description of the thermodynamics for a hard sphere fluid in a hard sphere matrix up to the close packing conditions~\cite{Hol12,Hol13,Hol14,Kalyuzh14}. Apart from the BH theory, we present the results obtained in the HTA approximation. Using the developed approach, the phase diagrams for a confined Lennard-Jones are built.
 Different matrix porosities and matrix particle sizes are considered. To check the accuracy of the approaches proposed in this paper,
a comparison with the MSA results as well as with the results of Monte Carlo simulations is performed.

\section{Theory}
\subsection{Reference system: HS fluid in HS matrix}

We start our theoretical consideration with the description of the reference
system. For this purpose we briefly recapitulate the main ideas of the SPT theory and present here the
expressions for the chemical potential and pressure of a HS fluid in a HS matrix, which
were obtained in our previous papers~\cite{Pat11,Hol12}, and which are needed in the current study.
The key point of the SPT theory consists in a derivation of the excess chemical potential of
an additional scaled particle of a variable size inserted in a fluid. This excess
chemical potential is equal to a work needed to create a cavity in a fluid
which is free from any other particles. For a small scaled particle in a HS
fluid in the presence of a porous medium, the expression for the excess chemical
potential is equal to \cite{Pat11}:
\begin{equation}
\label{hol2.1}
\beta\mu_\textrm{s}^\textrm{ex}=\ln p_{0}(\lambda_\textrm{s})-\ln\left[1-\eta_{1}\frac{(1+\lambda_\textrm{s})^{3}}{p_{0}(\lambda_\textrm{s})}\right],
\end{equation}
where $\beta=1/(k_\textrm{B} T)$, $k_\textrm{B}$ is the Boltzmann constant, $T$ is the temperature,
$\eta_{1}=\frac{1}{6}\pi\rho_{1}\sigma_{1}^{3}$ is the fluid packing
fraction, $\rho_{1}$ is the fluid density, $\sigma_{1}$ is the diameter of
HS fluid particles.
The term $p_{0}(\lambda_\textrm{s})=\exp(-\beta\mu_\textrm{s}^{0})$ is defined by the
excess chemical potential of the scaled particle confined in an empty matrix,
$\mu_\textrm{s}^{0}$. It has the meaning of probability to find  a cavity created by
the scaled particle in the matrix in the absence of fluid particles. We should note that
here we use conventional notations~\cite{Given92,Rosinberg99,Pizio2000,Trokh96,Trokh97,Hol09,Chen10,Pat11,Hol12,Hol13},
where the index ``1'' is used to denote a fluid component, the index ``0''
denotes matrix particles, while for the scaled particles the index ``s'' is used.

For a large scaled particle, the excess chemical potential is presented by the thermodynamic expression for the work needed to create a
macroscopic cavity inside the fluid, which at the same time is confined in a porous medium, and the corresponding expression can be presented as follows:
\begin{equation}
\label{hol2.2}
\beta\mu_\textrm{s}^\textrm{ex}=w(\lambda_\textrm{s})
+\beta\frac{PV_\textrm{s}}{p_{0}(\lambda_\textrm{s})}\,,
\end{equation}
where $P$ is the pressure of fluid, $V_\textrm{s}$ is the volume of a scaled particle.
The multiplier $1/p_{0}(\lambda_\textrm{s})$ appears due to an excluded volume occupied
by matrix particles. In this context, it should be mentioned that the probability $p_{0}(\lambda)$ is directly related to two different types of porosity \cite{Pat11,Hol12,Hol13}. The  first one
corresponds to the case of $\lambda_\textrm{s}=0$ and provides the geometrical porosity
\begin{equation}
\label{hol2.3}
\phi_{0}=p_{0}(\lambda_\textrm{s}=0),
\end{equation}
which depends only on the structure of a matrix and it is equal to the volume fraction of a void between the matrix particles.
For a HS fluid in a HS matrix, it is equal to
\begin{equation}
\label{hol2.4}
\phi_{0}=1-\eta_{0}\,,
\end{equation}
where $\eta_{0}=\frac16\pi \sigma_{0}^{3}\rho_{0}$,
$\rho_{0}=\frac{N_{0}}{V}$, $N_{0}$ is the number of matrix particles,
$\sigma_{0}$ is the diameter of the matrix particles, $V$ is the volume of the
system.

The second type of porosity corresponds to the case $\lambda_\textrm{s}=1$ and provides the probe particle porosity~\cite{Pat11,Hol12,Hol13}
\begin{equation}
\label{hol2.5}
\phi=p_{0}(\lambda_\textrm{s}=1)=\re^{-\beta \mu_1^0},
\end{equation}
which is defined by the excess chemical potential of fluid particles in the
limit of infinite dilution $\mu_{1}^{0}$. Using the SPT theory
\cite{Reiss59,Reiss60} for the case of a HS fluid in
a HS matrix, the following expression for $\phi$ can be derived:
\begin{equation}
\label{hol2.6}
\phi=(1-\eta_{0})\exp\left\{-\left[\frac{3\eta_{0}\tau}{1-\eta_{0}}+
\frac{3\eta_{0}\left(1+\frac12\eta_{0}\right)\tau^{2}}{(1-\eta_{0})^{2}}+
\frac{\beta P_{0}\eta_{0}}{\rho_{0}}\right]\tau^{3}\right\},
\end{equation}
where $\tau=\sigma_{1}/\sigma_{0}$, and $P_0$ is the bulk pressure of the matrix particles.
\begin{equation}
\label{hol2.7}
\frac{\beta P_{0}}{\rho_{0}}=\frac{(1+\eta_{0}+\eta_{0}^{2})}{(1-\eta_{0})^{3}}\,.
\end{equation}
According to the ansatz of SPT~\cite{Hol09,Chen10,Pat11,Hol12,Hol13,Reiss59,Reiss60}, $w(\lambda_\textrm{s})$ can be presented in the form of an expansion:
\begin{equation}
\label{hol2.8}
w(\lambda_\textrm{s})=w_{0}+w_{1}\lambda_\textrm{s}+\frac12w_{2}\lambda_\textrm{s}^{2}\,.
\end{equation}
Coefficients of this expansion can be found from the continuity of $\mu_\textrm{s}^\textrm{ex}$ and the corresponding
derivatives $\partial\mu_\textrm{s}/\partial\lambda_\textrm{s}$ and $\partial^{2}\mu_\textrm{s}/\partial\lambda_\textrm{s}^{2}$
at $\lambda_\textrm{s}=0$. After setting $\lambda_\textrm{s}=1$, the expression (\ref{hol2.2}) yields the relation between
the pressure $P$ and the excess chemical potential $\mu_{1}^\textrm{ex}$ of a fluid:
\begin{equation}
\label{hol2.9}
\beta(\mu_{1}^\textrm{ex}-\mu_{1}^{0})=-\ln\left(1-\eta_{1}/\phi_{0}\right)+A\frac{\eta_{1}/\phi_{0}}{1-\eta_{1}/\phi_{0}}
+B\frac{(\eta_{1}/\phi_{0})^{2}}{(1-\eta_{1}/\phi_{0})^{2}}+\frac{\beta P}{\phi}\frac{\eta_{1}}{\rho_{1}}\,,
\end{equation}
where the coefficients $A$ and $B$ determine the porous medium structure, and for a HS fluid in a HS matrix, they are as follows:
\begin{align}
A&=6+\frac{3\eta_{0}\tau(\tau+4)}{1-\eta_{0}}
+\frac{9\eta_{0}^{2}\tau^{2}}{(1-\eta_{0})^{2}}\,,\nonumber\\
B&=\frac92\left(1+\frac{\tau\eta_{0}}{1-\eta_{0}}\right)^{2}.
\label{hol2.10}
\end{align}
Using the Gibbs-Duhem equation, which relates the pressure of a fluid with
its total chemical potential $\mu_1=\ln(\Lambda_1^3 \rho_1)+\mu_1^\textrm{ex}$ as
\begin{equation}
\label{hol2.11}
\left(\frac{\partial P}{\partial\rho_{1}}\right)_{T}=\rho_{1}\left(\frac{\partial \mu_{1}}{\partial\rho_{1}}\right)_{T}
\end{equation}
one derives the fluid compressibility as
\begin{align}
\beta\left(\frac{\partial P}{\partial\rho_{1}}\right)_{T}&=\frac{1}{\left(1-\eta_{1}/\phi\right)}+(1+A)\frac{\eta_{1}/\phi_{0}}
{\left({1-\eta_{1}/\phi}\right)
\left(1-\eta_{1}/\phi_{0}\right)}\nonumber\\
&+(A+2B)\frac{\left(\eta_{1}/\phi_{0}\right)^{2}}{\left(1-\eta_{1}/\phi\right)\left(1-\eta_{1}/\phi_{0}\right)^{2}}\nonumber\\
&+
2B\frac{\left(\eta_{1}/\phi_{0}\right)^{3}}{\left(1-\eta_{1}/\phi\right)\left(1-\eta_{1}/\phi_{0}\right)^{3}}\,.
\label{hol2.12}
\end{align}
After dividing the expression (\ref{hol2.12}) by $\rho_1$ and subsequently integrating it over $\rho_1$,
one obtains the chemical potential:
\begin{align}
\beta \mu_{1}^\textrm{SPT2}&=\ln(\Lambda_1^3 \rho_1)-\ln(\phi)-\ln(1-\eta_{1}/\phi)+(A+1)\frac{\phi}{\phi-\phi_{0}}\ln\frac{1-\eta_{1}/\phi}{1-\eta_{1}/\phi_{0}}\nonumber\\
&+(A+2B)\frac{\phi}{\phi-\phi_{0}}\left(\frac{\eta_{1}/\phi_{0}}{1-\eta_{1}/\phi_{0}}-\frac{\phi}{\phi-\phi_{0}}\ln\frac{1-\eta_{1}/\phi}
{1-\eta_{1}/\phi_{0}}\right)\nonumber\\
&+2B\frac{\phi}{\phi-\phi_{0}}\left[\frac12\frac{(\eta_{1}/\phi_{0})^{2}}{(1-\eta_{1}/\phi_{0})^{2}}-\frac{\phi}{\phi-\phi_{0}}
\frac{\eta_{1}/\phi_{0}}{1-\eta_{1}/\phi_{0}}\right.\nonumber\\
&\left.+\frac{\phi^{2}}{(\phi-\phi_{0})^{2}}
\ln\frac{1-\eta_{1}/\phi} {1-\eta_{1}/\phi_{0}}\right].
\label{hol2.13}
\end{align}
It is worth noting that the second term $-\ln(\phi)$ in (\ref{hol2.13}) follows from the relation~(\ref{hol2.5}) and
the corresponding substitution $\beta \mu_1^0=-\ln(\phi)$.
Similarly, integration of the right-hand side of expression~(\ref{hol2.12}) over $\rho_1$ leads to the pressure
\begin{align}
\left(\frac{\beta P}{\rho_{1}}\right)^\textrm{SPT2}&=-\frac{\phi}{\eta_{1}}\ln\frac{1-\eta_{1}/\phi}{1-\eta_{1}/\phi_{0}}+
(1+A)\frac{\phi}{\eta_{1}}\frac{\phi}{\phi-\phi_{0}}\ln\frac{1-\eta_{1}/\phi}{1-\eta_{1}/\phi_{0}}\nonumber\\
&+(A+2B)\frac{\phi}{\phi-\phi_{0}}\left(\frac{1}{1-\eta_{1}/\phi_{0}}-\frac{\phi}{\eta_{1}}\frac{\phi}{\phi-\phi_{0}} \ln\frac{1-\eta_{1}/\phi}
{1-\eta_{1}/\phi_{0}}\right)\nonumber\\
&+2B\frac{\phi}{\phi-\phi_{0}}\left[\frac12\frac{\eta_{1}/\phi_{0}}{(1-\eta_{1}/\phi_{0})^{2}}-\frac{2\phi-\phi_{0}}{\phi-\phi_{0}}\frac{1}{1-\eta_{1
}/\phi_{0}}\right.\nonumber\\
&\left.+
\frac{\phi}{\eta_{1}}\frac{\phi^{2}}{(\phi-\phi_{0})^{2}}\ln\frac{1-\eta_{1}/\phi}{1-\eta_{1}/\phi_{0}}\right].
\label{hol2.14}
\end{align}
The expressions (\ref{hol2.13}) and (\ref{hol2.14}) are considered as the expression derived within the framework
of the SPT2 approach~\cite{Pat11}. A simple analysis of (\ref{hol2.13}) and (\ref{hol2.14}) shows
that they have two divergences at $\eta_{1}=\phi$ and $\eta_{1}=\phi_{0}$.
Since $\phi<\phi_{0}$, the divergence at $\eta_{1}=\phi$ occurs at lower densities. However, from the geometrical point
of view, this divergence should appear at higher densities near the maximum value of the fluid
packing fraction available for a fluid in a given matrix. Different corrections improving the SPT2 approach
were proposed in~\cite{Pat11,Hol12,Hol13}. Here, we consider two of them, which provide rather accurate
results in comparison with computer simulations. The first of them known as SPT2b can be derived
if $\phi$ is replaced by $\phi_{0}$ everywhere in~(\ref{hol2.12}) except for the first term.
In this case, the chemical potential and the pressure of a confined fluid are as follows:
\begin{align}
\beta \mu_{1}^\textrm{SPT2b}&=\ln(\Lambda_1^3 \rho_1)-\ln(\phi)-\ln(1-\eta_{1}/\phi)+(1+A)\frac{\eta_{1}/\phi_{0}}{1-\eta_{1}/\phi_{0}}\nonumber\\
&+\frac12
(A+2B)\frac{(\eta_{1}/\phi_0)^{2}}{(1-\eta_{1}/\phi_{0})^{2}}
+\frac{2}{3}B\frac{(\eta_{1}/\phi_{0})^{3}}{(1-\eta_{1}/\phi_{0})^{3}}\,,
\label{hol2.15}
\end{align}
\begin{align}
\left(\frac{\beta P}{\rho_{1}}\right)^\textrm{SPT2b}&=-\frac{\phi}{\eta_{1}}\ln\left(1-\frac{\eta_{1}}{\phi}\right)
+\frac{\phi_{0}}{\eta_{1}}\ln\left(1-\frac{\eta_{1}}{\phi_{0}}\right) +\frac{1}{1-\eta_{1}/{\phi_{0}}}\nonumber\\
&+\frac{A}{2}\frac{\eta_{1}/\phi_{0}}{(1-\eta_{1}/\phi_{0})^{2}}
+\frac{2B}{3}\frac{(\eta_{1}/\phi_{0})^{2}}{(1-\eta_{1}/\phi_{0})^{3}}\,.
\label{hol2.16}
\end{align}
The second approximation referred to as SPT2b1 can be derived from SPT2b by removing
the divergence at $\eta_{1}=\phi$ by an expansion of the logarithmic
term in~(\ref{hol2.15})
\begin{equation}
\label{hol2.17}
-\ln \left(1-\eta_{1}/\phi\right)\approx -\ln\left(1-\eta_{1}/\phi_{0}\right)
+\frac{\eta(\phi_{0}-\phi)}{\phi_{0}\phi(1-\eta_{1}/\phi_{0})}\,.
\end{equation}
As a consequence, one obtains the following expressions within the SPT2b1 approximation:
\begin{align}
\beta \mu_{1}^\textrm{SPT2b1}&=\ln(\Lambda_1^3 \rho_1)-\ln(\phi)-\ln(1-\eta_{1}/\phi_{0})
+(1+A)\frac{\eta_{1}/\phi_{0}}{1-\eta_{1}/\phi_{0}}\nonumber\\
&+\frac{\eta_{1}(\phi_{0}-\phi)}{\phi_{0}\phi(1-\eta_{1}/\phi_{0})}
+\frac12(A+2B)\frac{(\eta_{1}/\phi_{0})^{2}}{(1-\eta_{1}/\phi_{0})^{2}}+\frac{2}{3}B\frac{(\eta_{1}/\phi_{0})^{3}}
{(1-\eta_{1}/\phi_{0})^{3}}\,,
\label{hol2.18}
\end{align}
\begin{align}
\left(\frac{\beta P}{\rho_{1}}\right)^\textrm{SPT2b1}&=\frac{1}{1-\eta_{1}/\phi_{0}}\frac{\phi_{0}}{\phi}+\left(\frac{\phi_{0}}{\phi}-1\right)
\frac{\phi_{0}}{\eta_{1}}\ln\left(1-\frac{\eta_{1}}{\phi_{0}}\right)\nonumber\\
&+\frac{A}{2}\frac{\eta_{1}/\phi_{0}}{(1-\eta_{1}/\phi_{0})^{2}}
+\frac{2B}{3}\frac{(\eta_{1}/\phi_{0})^{2}}{(1-\eta_{1}/\phi_{0})^{3}}\,.
\label{hol2.19}
\end{align}

\subsection{BH perturbation theory for simple fluid in random porous medium}

The next step of theoretical treatment is connected with a consideration of
an attractive part of interaction. We consider a simple fluid with an intermolecular interaction in the form
\begin{eqnarray}
{\it v}_{11}(r)=\left\{\begin{array}{ll}
\infty, & \hbox{$r<\sigma_{1}$},\\
u_{11}(r),& \hbox{$r>\sigma_{1}$},
\end{array}\right.
\label{hol2.20}
\end{eqnarray}
where $u_{11}(r)\leqslant0$ is a pure attractive part of interaction.

In order to take into account the attractive part of interaction, in this subsection we generalize the BH perturbation theory for the
case of a fluid in a random HS matrix. To this end, we use the replica trick~\cite{Given92} according to which
a system of a fluid in a matrix of unmovable (frozen) particles can be replaced by an equilibrium mixture consisting of the movable (annealed)
matrix particles and $s$ identical copies (or replicas) of a fluid. The condition is also set that the fluid replicas from different copies do not interact with each other, but they interact with the matrix.
Such a system can be described in a standard way using the liquid state theories, and the properties of a fluid can be obtained
by considering the limit $s\rightarrow0$.
Therefore, the Helmholtz free energy of a fluid in a matrix can be presented as~\cite{Kierlik97}:
\begin{equation}
\label{hol2.21}
F=\lim_{s\rightarrow0}\frac{\rm d}{{\rm d}s}F(s),
\end{equation}
where $F(s)$ is the free energy of the $(s+1)$-component equilibrium mixture.

Within the framework of the Zwanzig high-temperature perturbation theory~\cite{Zwanzig54}, the first term of
the free energy expansion corresponds to the high-temperature approximation:
\begin{equation}
\label{hol2.22}
\frac{\beta(F-F_{0})^\textrm{HTA}}{V}=\frac12\rho_{1}^{2}\beta\int{\rm d}\bar{r}g_{11}^\textrm{HS}(r)u_{11}(r),
\end{equation}
which is of the same form as in the optimized cluster expansions~\cite{Kierlik97}. $F_0$ and $g_{11}^\textrm{HS}(r)$ are the free energy
and the pair distribution function of a HS fluid in a HS matrix, respectively.

The second  correction term of Zwanzig expansion involves the three- and
four-body distribution functions, for which it is difficult to find simple
satisfactory approximations. Therefore, instead of this, we follow Barker and
Henderson~\cite{BarkerHen67,Barker76} and we write the free energy in the form:
\begin{align}
\frac{\beta(F-F_{0})^\textrm{BH}}{V}&=\frac12\rho_{1}^{2}\beta\int{\rm d}\bar{r}g_{11}^\textrm{HS}(r)u_{11}(r)\nonumber\\
&-\frac14\rho_{1}^{2}\beta \left(\frac{\partial \rho_{1}}
{\partial P}\right)_{T}^\textrm{HS}\int{\rm d}\bar{r}
u_{11}^{2}(r)\left(\frac{\partial\left[\rho_{1}g_{11}^\textrm{HS}(r)\right]}{\partial \rho_{1}}\right)_\textrm{HS},
\label{hol2.23}
\end{align}
where we use the same semimacroscopic arguments as in~\cite{BarkerHen67,Barker76} and neglect the cross-correlation
terms between fluids from different replicas.

Differentiating the expressions (\ref{hol2.22}) and (\ref{hol2.23}) with
respect to the fluid density, one derives the expression for the chemical
potential of a fluid:
\begin{equation}
\label{hol2.24}
\beta\mu_{1}=\beta\mu_{1}^\textrm{HS}+\beta\mu_{1}^\textrm{HTA}+\beta\mu_{1}^\textrm{BH},
\end{equation}
where $\mu_{1}^\textrm{HS}$ is the HS contribution of the reference system, which can be given by equation~(\ref{hol2.15}) within the SPT2b approximation
or by the equation~(\ref{hol2.18}) within the SPT2b1 approximation.
The first term of (\ref{hol2.23}) corresponds to the HTA approximation, and it has the following form:
\begin{equation}
\label{hol2.25}
\beta\mu_{1}^\textrm{HTA}=2\pi\beta\rho_{1}\left[2 I(\rho_{1})+\rho_{1}\frac{\partial}{\partial\rho_{1}}I(\rho_{1})\right].
\end{equation}
The contribution coming from the BH approximation as the second term of~(\ref{hol2.23}) is as follows:
\begin{align}
\beta\mu_{1}^\textrm{BH}&=-\pi\beta\rho_{1}\left[2J(\rho_{1})\left(\frac{\partial \rho_{1}}{\partial P}\right)^\textrm{HS}_{T}+
\rho_{1}J(\rho_{1})\frac{\partial}{\partial\rho_{1}}\left(\frac{\partial\rho_{1}}{\partial P}\right)^\textrm{HS}_{T}\right.\nonumber\\
&+4\rho_{1}\left(\frac{\partial\rho_{1}}{\partial P}\right)^\textrm{HS}_{T}\frac{\partial}{\partial\rho_{1}}J(\rho_{1})+\rho_{1}^{2}\frac{\partial J(\rho_{1})}{\partial\rho_{1}}
\frac{\partial}{\partial\rho_{1}}\left(\frac{\partial\rho_{1}}{\partial P}\right)^\textrm{HS}_{T}\nonumber\\
&\left.+\rho_{1}^{2}\left(\frac{\partial\rho_{1}}{\partial P}\right)^\textrm{HS}_{T}\frac{\partial^{2}J(\rho_{1})}{\partial\rho_{1}^{2}}\right].
\label{hol2.26}
\end{align}
The expressions for the first and second derivatives of the isothermal compressibility
with respect to the fluid density for the reference system,
$\left(\frac{\partial\rho_{1}}{\partial P}\right)^\textrm{HS}_{T}$ and
$\frac{\partial}{\partial\rho_{1}}\left(\frac{\partial\rho_{1}}{\partial
P}\right)^\textrm{HS}_{T}$, are found using equations (\ref{hol2.16})
and (\ref{hol2.19}) within the framework of the SPT2b or SPT2b1
approximations, correspondingly, and they are presented in Appendix.
The functions $I(\rho_1)$ and $J(\rho_1)$ are the integrals from the first and second terms
of (\ref{hol2.23})
\begin{align}
I(\rho_{1})=\int_{0}^{\infty}g_{11}^\textrm{HS}(r)u_{11}(r)r^{2}{\rm d}r,\nonumber\\
J(\rho_{1})=\int_{0}^{\infty}g_{11}^\textrm{HS}(r)u_{11}^{2}(r)r^{2}{\rm d}r.
\label{hol2.27}
\end{align}
These integrals contain the pair distribution function $g_{11}^\textrm{HS}(r)$, which is unknown for this moment and will be considered
separately in the next subsection.

The pressure can be calculated by differentiating the expression
(\ref{hol2.23}) with respect to the volume of a system or from the general
thermodynamical relation:
\begin{equation}
\label{hol2.28}
\beta P=\beta\rho_{1}\mu_{1}-\beta\frac{ F}{V}\,.
\end{equation}
The general form of the pressure is as follows:
\begin{equation}
\label{hol2.29}
\beta P=\beta P^\textrm{HS}+\beta P^\textrm{HTA}+\beta P^\textrm{BH},
\end{equation}
where $P^\textrm{HS}$ is the HS contribution given by (\ref{hol2.16}) within the framework of SPT2b approximation or by (\ref{hol2.19})
within the framework of SPT2b1 approximation.
The HTA term for the pressure is as follows:
\begin{equation}
\label{hol2.30}
\frac{\beta P^\textrm{HTA}}{\rho_{1}}=2\pi\beta\rho_{1}\left[I(\rho_{1})+\rho_{1}\frac{\partial}{\partial\rho_{1}}I(\rho_{1})\right].
\end{equation}
The contribution of BH term is as follows:
\begin{align}
\frac{\beta P^\textrm{BH}}{\rho_{1}}&=-\pi\beta\rho_{1}\left[J(\rho_{1})\left(\frac{\partial\rho_{1}}{\partial P}\right)^\textrm{HS}_{T}+\rho_{1}
J(\rho_{1})\frac{\partial}{\partial\rho_{1}}\left(\frac{\partial\rho_{1}}{\partial P}\right)^\textrm{HS}_{T}\right.\nonumber\\
&+3\rho_{1}\left(\frac{\partial\rho_{1}}{\partial P}\right)^\textrm{HS}_{T}\frac{\partial}{\partial\rho_{1}}J(\rho_{1})+\rho_{1}^{2}
\frac{\partial J(\rho_{1})}{\partial\rho_{1}}\frac{\partial}{\partial\rho_{1}}\left(\frac{\partial\rho_{1}}{\partial P}\right)^\textrm{HS}_{T}\nonumber\\
&\left.+\rho_{1}^{2}\left(\frac{\partial\rho_{1}}{\partial P}\right)^\textrm{HS}_{T}\frac{\partial^{2}J(\rho_{1})}{\partial\rho_{1}^{2}}\right].
\label{hol2.31}
\end{align}

\subsection{The replica Ornstein-Zernike equations}

The pair distribution function $g_{11}^\textrm{HS}(r)$ needed to calculate the integrals~(\ref{hol2.27}) can be obtained
from a solution of the so-called replica Ornstein-Zernike~(ROZ) equations, which were derived by Given and Stell~\cite{Given92}
using the replica trick:
\begin{align}
h_{00}&=c_{00}+\rho_{0}c_{00}\otimes h_{00}\,,\nonumber\\
h_{10}&=c_{10}+\rho_{0}c_{10}\otimes h_{00}+\rho_{1}c_\textrm{c}\otimes h_{10}\,,\nonumber\\
h_{11}&=c_{11}+\rho_{0}c_{10}\otimes h_{01}+\rho_{1}c_\textrm{c}\otimes h_{11}+\rho_{1}c_\textrm{b}\otimes h_\textrm{c}\,,\nonumber\\
h_\textrm{c}&=c_\textrm{c}+\rho_{1}c_\textrm{c}\otimes h_\textrm{c}\,,
\label{hol2.32}
\end{align}
where the symbol $\otimes$ denotes a convolution.
The pair and direct fluid-fluid correlation functions are separated into the connected and blocked parts
\begin{align}
&h_{11}(r)=h^\textrm{c}(r)+h^\textrm{b}(r)\nonumber,\\
&c_{11}(r)=c^\textrm{c}(r)+c^\textrm{b}(r).
\label{hol2.33}
\end{align}
As usual in the liquid state theory~\cite{Yukhnovski80,Hansen06}, the ROZ equations need
additional closure relations. The Percus-Yevick~(PY) approximation is used for a HS fluid in a HS matrix considered in
this paper as the reference system.
As it was shown in \cite{Vega93,Lomba93}, this approximation gives results
for the pair distribution functions which are in good agreement with computer
simulations. In the PY approximation for a HS fluid in a HS matrix,
the blocking direct correlation function is zero $c_\textrm{b}(r)=0$, thus $c_\textrm{c}(r)=c_{11}(r)$.
Therefore, the closure conditions in our case are as follows:
\begin{eqnarray}\begin{array}{llll}
h_{00}(r)=-1 &\text{if}\quad r<\sigma_{0}\,,  &c_{00}(r)=0&\text{if} \quad r>\sigma_{0}\,,\\
h_{10}(r)=-1 &\text{if}\quad r<\sigma_{01}\,, &c_{10}(r)=0&\text{if} \quad r>\sigma_{01}\,,\\
h_{11}(r)=-1 &\text{if}\quad r<\sigma_{1}\,,  &c_{11}(r)=0&\text{if} \quad r>\sigma_{1}\,,
\end{array}
\label{hol2.34}
\end{eqnarray}
where $\sigma_{01}=\frac12(\sigma_{0}+\sigma_{1})$.

\begin{figure}[!t]
\centerline{
\includegraphics [width=0.48\textwidth]  {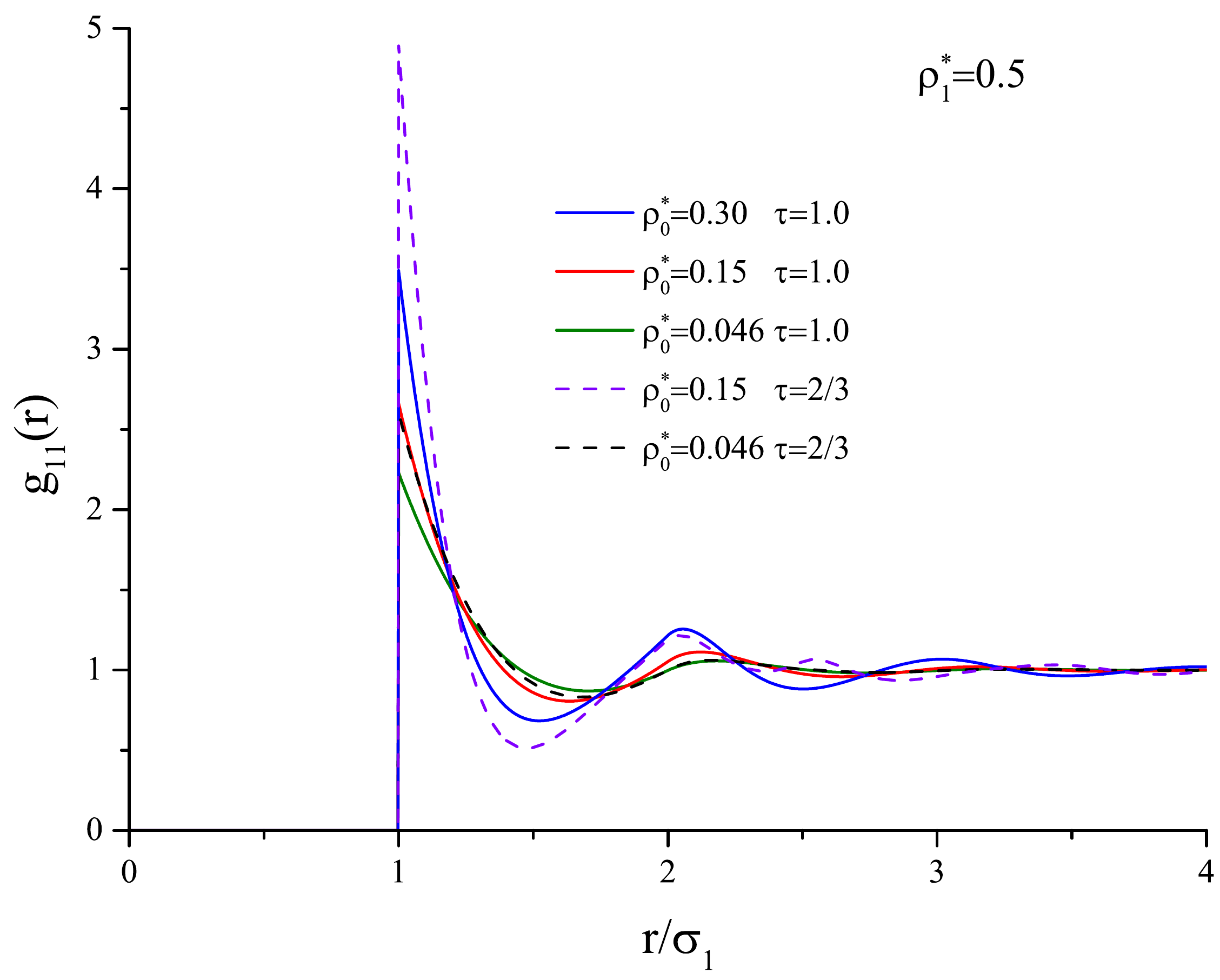}
\includegraphics [width=0.48\textwidth]  {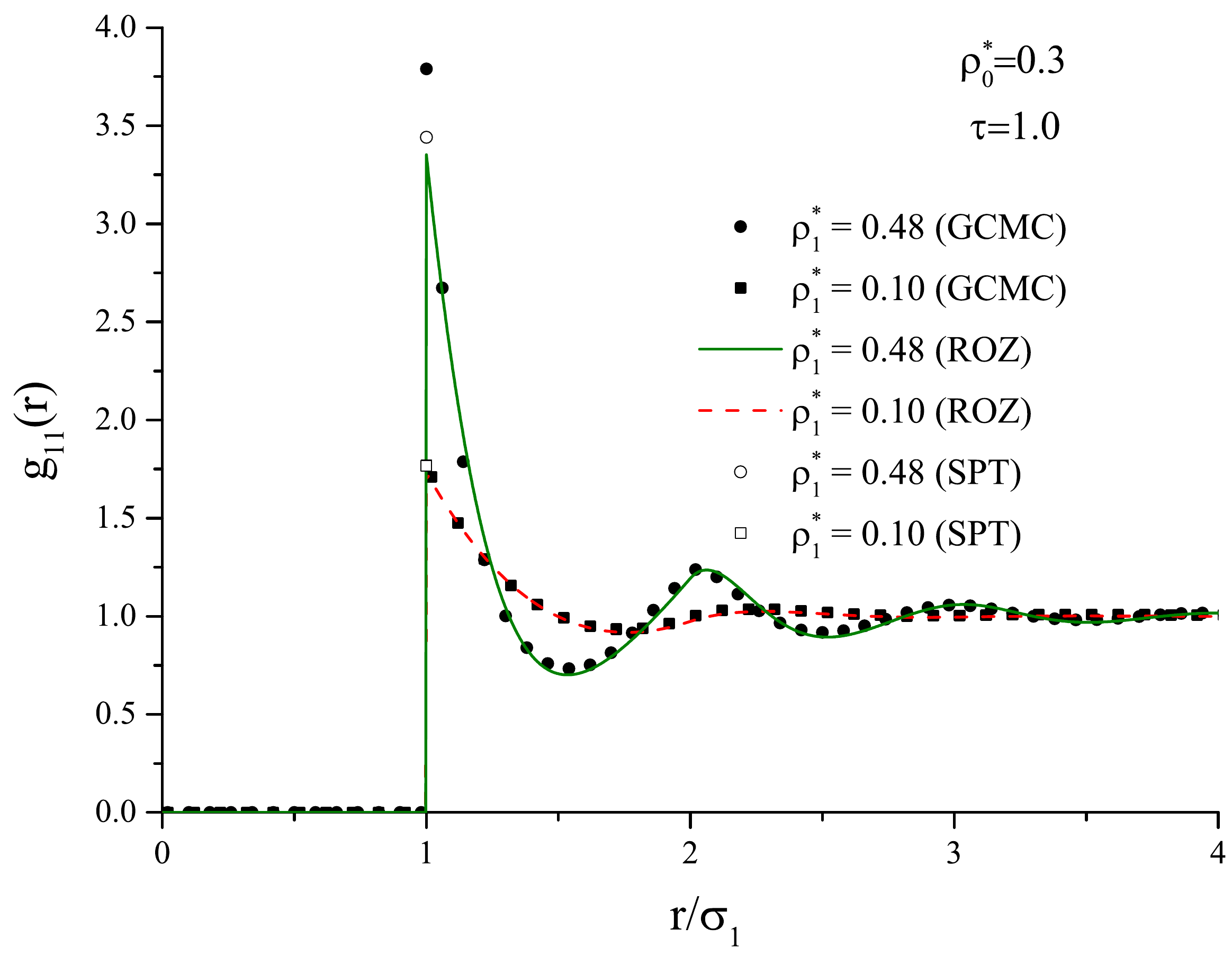}
}
\caption{(Color online) Fluid-fluid pair distribution functions $g^\textrm{HS}_{11}(r)$ of a HS fluid in a HS matrix for different parameters. Left-hand panel: the different matrix densities $\rho_{0}^{*}=\rho_{0}\sigma_{1}^{3}$
and the size ratios of fluid and matrix particles $\tau=\sigma_1/\sigma_0$,
but the fixed fluid density $\rho_{1}^{*}=\rho_{1}\sigma_{1}^{3}=0.5$.
Right-hand panel: the different fluid densities $\rho_{1}^{*}$,
but the fixed matrix density $\rho_{0}^{*}=0.3$.}
\label{Fig1}
\end{figure}

The set of equations~(\ref{hol2.32}) in combination with the closure relations~(\ref{hol2.34}) is solved
numerically using the hybrid Newton-Raphson procedure~\cite{Labik85}.
Some of the results for $g_{11}^\textrm{HS}(r)=1+h_{11}(r)$ are presented in
figure~\ref{Fig1}. As one can see, the function $g_{11}^\textrm{HS}(r)$ shows a typical
behaviour for a HS fluid. In the left-hand panel of figure~\ref{Fig1} it is shown that for the fixed fluid density,
an increase of the matrix density, i.e., lowering the matrix porosity, leads to a short-range order increase.
The same effect is observed if the fluid density is increased, but the matrix density is kept constant~(figure~\ref{Fig1}, right-hand panel).
There is also a comparison of the ROZ results with the Monte-Carlo simulations for the low and high fluid densities of a HS fluid in a HS matrix.
It is clearly seen that they fit very well, except the contact value $g_{11}(\sigma_1^+)$ for the dense fluid, which is somewhat lower
in the case of ROZ equations. Besides, there are two points of the contact value obtained from the SPT for a fluid in a matrix~\cite{Kalyuzh14},
which are a bit higher than the values of ROZ, thus closer to the corresponding simulation results.
In this paper, for a comparison we also consider the mean spherical
approximation~(MSA) for the studied model of a fluid with the pair interaction between particles
in the form~(\ref{hol2.20}). Similarly to the HTA and BH approximations, to calculate the chemical potential and the pressure within
the MSA approximation, the correlation functions are needed. For this purpose, we can also use the ROZ equations in combination
with the corresponding closure relations:
\begin{eqnarray}\begin{array}{llll}
h_{00}(r)=-1 &\text{if}\quad r<\sigma_{0}\,, &c_{00}(r)=0&\text{if}\quad r>\sigma_{0}\,,\\
h_{10}(r)=-1 &\text{if}\quad r<\sigma_{01}\,,&c_{10}(r)=0&\text{if} \quad r>\sigma_{01}\,,\\
h_{11}(r)=-1 &\text{if}\quad r<\sigma_{1}\,, &c_{11}(r)=c_\textrm{c}(r)=-\beta u_{11}(r)&\text{if}\quad r>\sigma_{1}\,.
\end{array}
\label{hol2.35}
\end{eqnarray}
Again, using the replica procedure~\cite{Given92} and the general expressions of Hoye and
Stell~\cite{Hoye77} for thermodynamic properties of an equilibrium mixture of $(s+1)$ components in
the limit $s\rightarrow 0$, one obtains the chemical potential and pressure of
a fluid within the framework of the MSA approximation:
\begin{align}
&\beta\mu_{1}=\beta\mu_{1}^\textrm{HS}+\beta\mu_{1}^\textrm{MSA}\nonumber\\
&\beta P=\beta P^\textrm{HS}+\beta P^\textrm{MSA},
\label{hol2.36}
\end{align}
where ``HS'' denotes the reference system, i.e., the HS system. For the perturbation part within the MSA approximation,
the expressions for the chemical potential is~\cite{Kierlik97}:
\begin{align}
&\beta\mu_{1}^\textrm{MSA}=2\pi\beta\rho_{1}\int_{0}^{\infty}u_{11}(r)g_{11}(r)r^{2}{\rm d}r\nonumber\\
&-2\pi\rho_{1}\int_{0}^{\infty}\left[c_{11}(r)-c_{11}^\textrm{HS}(r)\right]r^{2}{\rm d}r
-2\pi\rho_{0}\int_{0}^{\infty}\left[c_{01}(r)-c_{01}^\textrm{HS}(r)\right]r^{2}{\rm d}r.
\label{hol2.37}
\end{align}
In the same way the pressure of a fluid confined in a matrix can be found:
\begin{align}
\beta P^\textrm{MSA}/\rho_{1}&=\frac13\pi\rho_{1}\sigma_{1}^{3}
\left\{g_{11}^{2}(\sigma_{1}^{+})-\left[g_{11}^{2}(\sigma_{1}^{+})\right]_\textrm{HS}\right\}\nonumber\\
&+\frac13\pi\rho_{0}\sigma_{10}^{3}\left\{g_{10}^{2}(\sigma_{10}^{+})-\left[g_{10}^{2}(\sigma_{10}^{+})\right]_\textrm{HS}\right\}\nonumber\\
&+\frac{2}{3}\pi\rho_{1}\int_{0}^{\infty}g_{11}(r)\frac{\partial u_{11}(r)}{\partial r}r^{3}{\rm d}r,
\label{hol2.37a}
\end{align}
where $c_{11}^\textrm{HS}(r)$ and $c_{10}^\textrm{HS}(r)$ are the fluid-fluid and
fluid-matrix direct correlation functions for a HS fluid in a HS matrix,
$g_{11}(\sigma_{1}^{+})$ and $g_{10}(\sigma_{10}^{+})$ are the contact
values of pair distribution functions $g_{11}(r)=1+h_{11}(r)$ and
$g_{10}(r)=1+h_{10}(r)$.

\subsection{Some calculation details}

In the present study we consider a fluid of spherical molecules confined in a HS matrix. According to~(\ref{hol2.20}), the fluid--fluid interaction
is decomposed into a hard-sphere part and a Lennard-Jones~(12-6) tail following Weeks, Chandler and Andersen~\cite{Weeks71}, i.e.,
\begin{eqnarray}
u_{11}(r)=\left\{\begin{array}{ll}
-\epsilon,\,\,&\sigma_{1}< r<\sqrt[6]{2}\sigma_{1}\,,\\
4\epsilon\left[\left(\sigma_{1}/r\right)^{12}-\left(\sigma_{1}/r\right)^{6}\right],\,\, &r>\sqrt[6]{2}\sigma_{1}\,.
\end{array}\right.
\label{hol2.38}
\end{eqnarray}
\begin{figure}[!b]
\centerline{
\includegraphics [width=0.49\textwidth]  {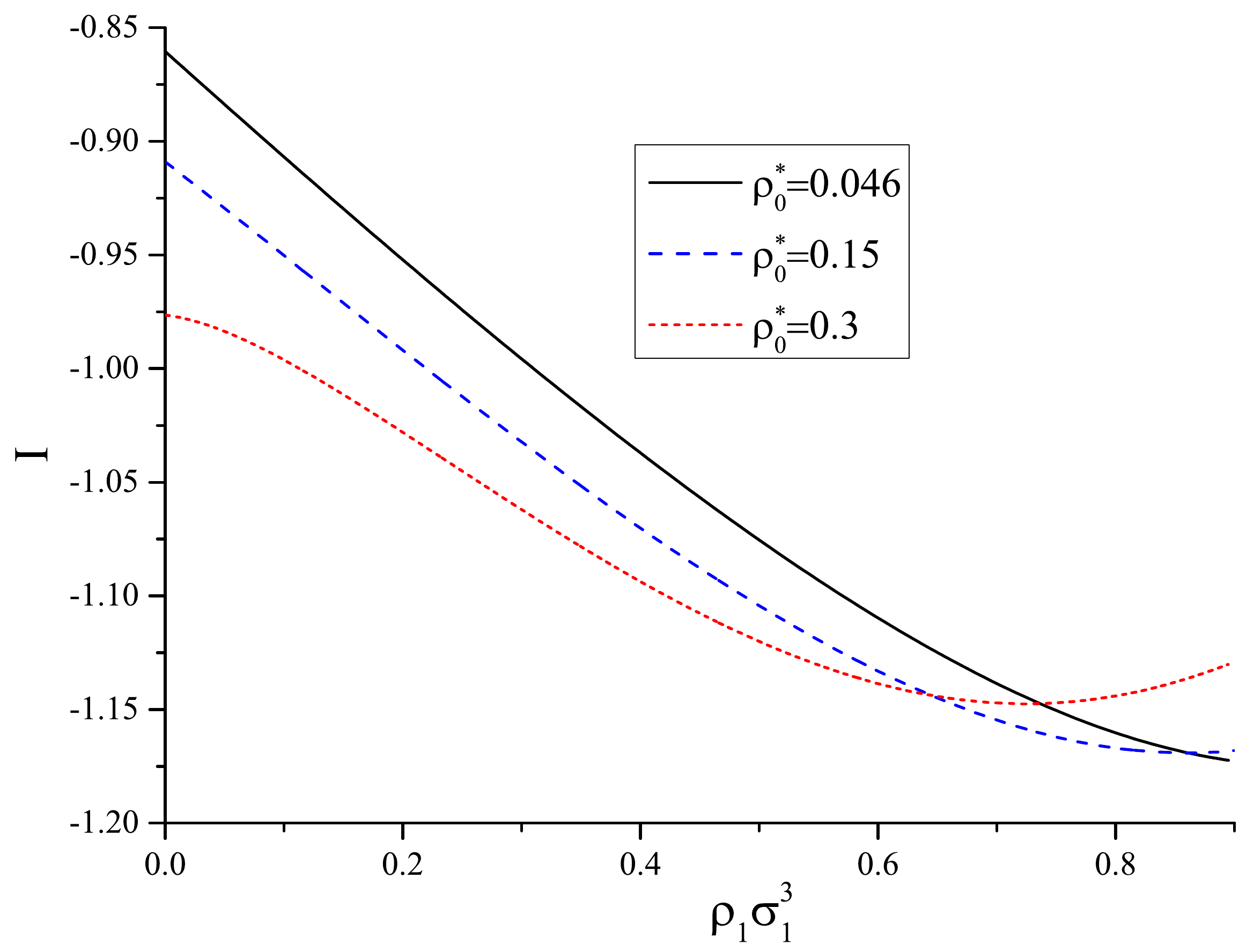}
\includegraphics [width=0.47\textwidth]  {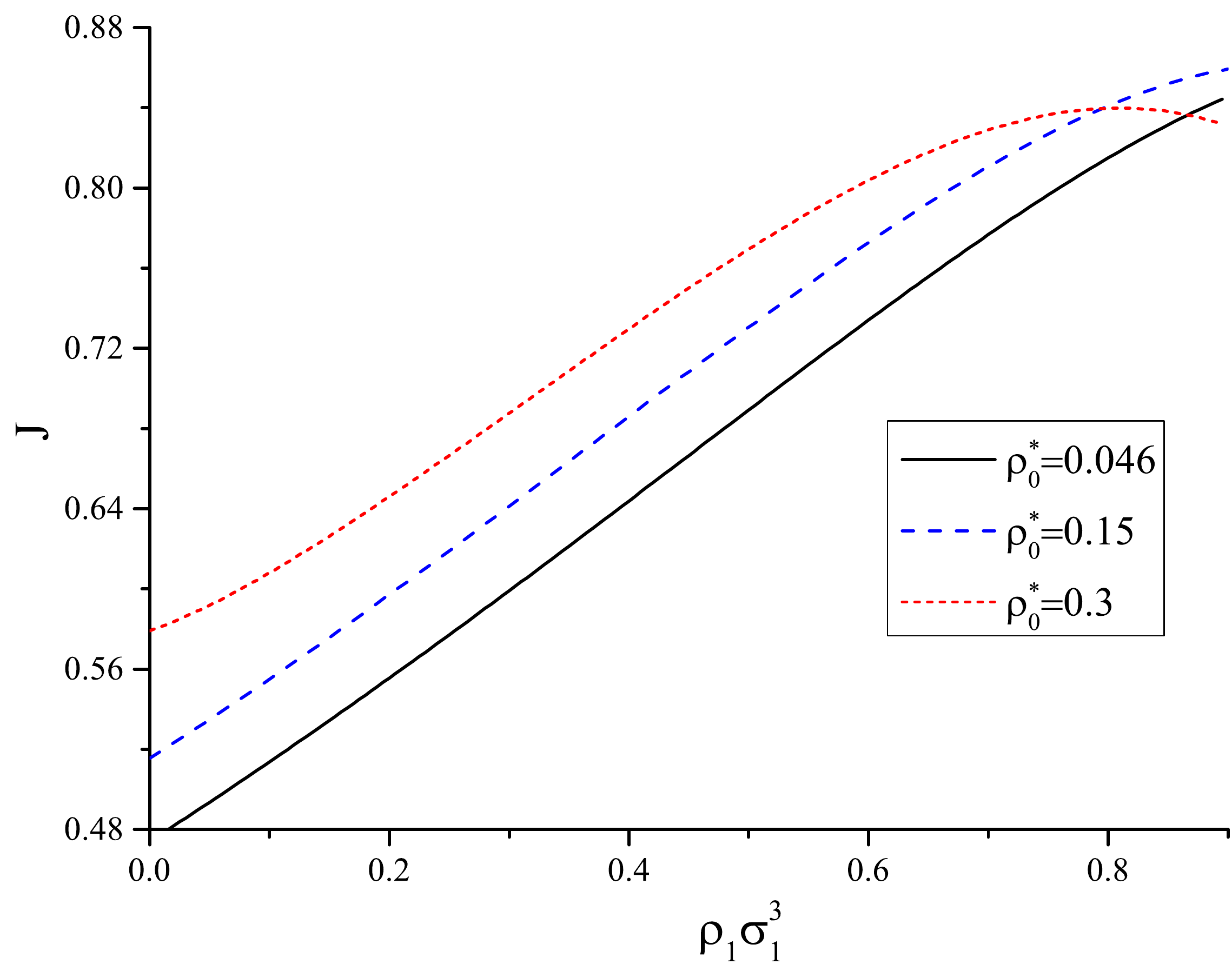}
}
\caption{(Color online) The integrals $I(\rho_{1})$ and $J(\rho_{1})$ defined by (\ref{hol2.27}) for different matrix densities.}
\label{Fig2}
\end{figure}
Applying a hard core potential for the repulsive potential, we
avoid difficulties connected with the temperature dependence of the fluid diameter, which
requires a special treatment in the case of soft repulsion~\cite{BarkerHen67}.
The Lennard-Jones attractive tail is truncated at $r=2.5\sigma_{1}$ in order to compare our results with the
corresponding computer simulations~\cite{Alvarez99}.

The functions $I(\rho_{1})$ and $J(\rho_{1})$ used in the generalized BH
theory should be calculated from the integrals~(\ref{hol2.27}). The dependencies of
$I(\rho_{1})$ and $J(\rho_{1})$ on the fluid density are illustrated in
figure~\ref{Fig2}. As one can see, these dependencies are rather smooth, that is why they can be
interpolated as polynomials. It is noticed that polynomials of the order of 9 provide a satisfactory fitting
for the given functions. Since finally the functions $I(\rho_{1})$ and $J(\rho_{1})$ are polynomials,
all first and second derivatives of these functions used in (\ref{hol2.25})--(\ref{hol2.26}) and (\ref{hol2.30})--(\ref{hol2.31})
are calculated analytically.

Having the chemical potential and pressure as functions of $\rho_1$ (or $\eta_1$) at different temperatures,
one can calculate the coexistence curves of the liquid--vapour phase transition.
For this purpose, we solve a set of two non-linear equations which follows from the conditions of thermodynamic equilibrium:
\begin{align}
\mu_{1}(\rho_{1}^\textrm{v}, T)&=\mu_{1}(\rho_{1}^\textrm{l}, T),\nonumber\\
P(\rho_{1}^\textrm{v}, T)&= P(\rho_{1}^\textrm{l}, T),
\label{hol2.39}
\end{align}
where $\rho_{1}^\textrm{v}$ and $\rho_{1}^\textrm{l}$ are the fluid density of
vapour and liquid phases, respectively. The numerical solution of these equations is realized
using the Newton-Raphson algorithm. Thus, the liquid--vapour phase diagrams of a fluid confined
in HS matrices are constructed.

\section{Results and discussions}

\begin{figure}[!b]
\centerline{
\includegraphics [width=0.5\textwidth]  {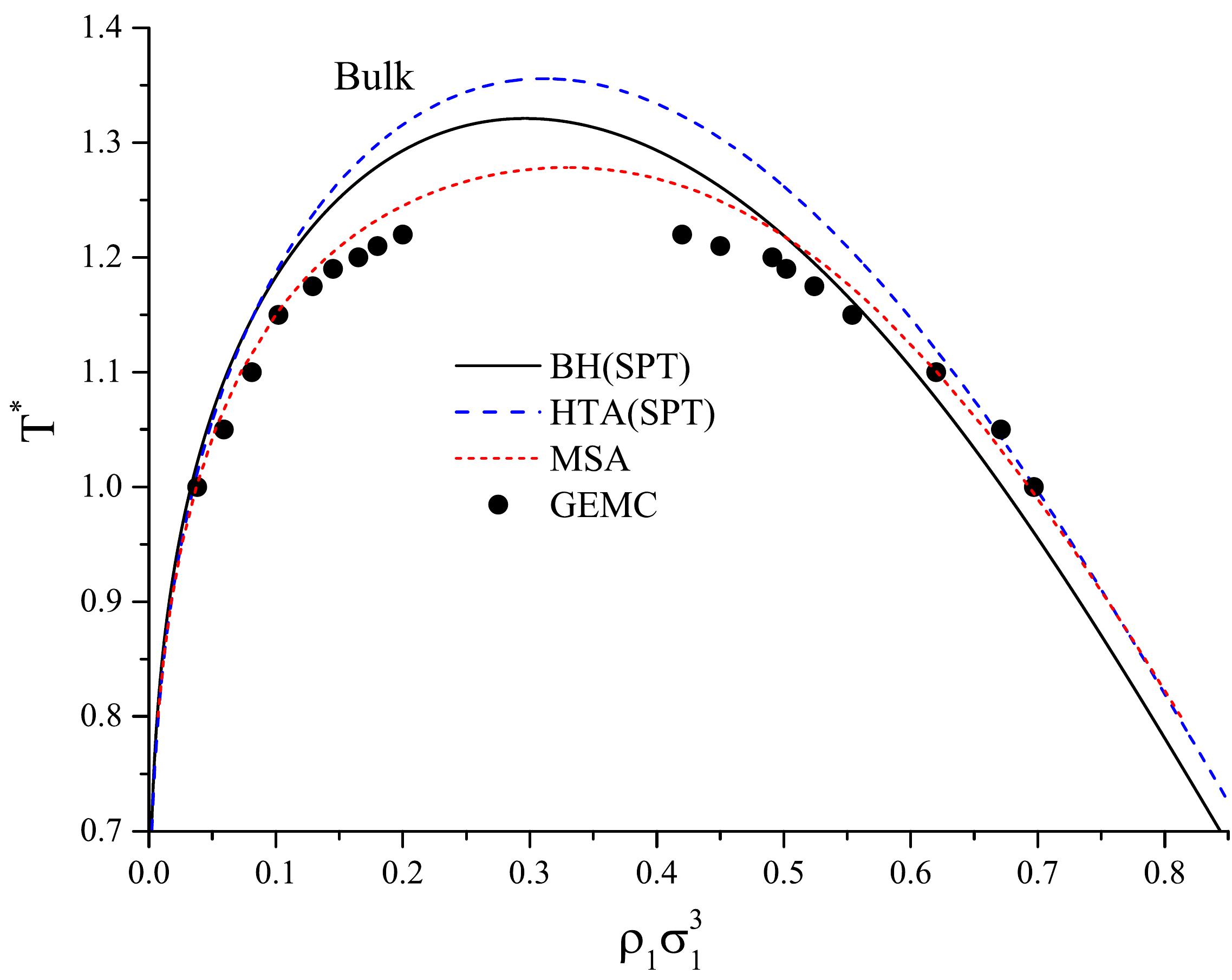}
}
\caption{(Color online) Liquid--vapour phase diagram for the bulk fluid ($\rho_{0}=0$) with the fluid-fluid interaction~(\ref{hol2.20}) and
(\ref{hol2.38}), where $T^{*}=k_\textrm{B}T/\epsilon$.
Solid line corresponds to BH theory, dotted line~--- HTA, dashed line~--- MSA, symbols~--- GEMC simulation results~\cite{Alvarez99}.}
\label{Fig3}
\end{figure}

We apply the theory presented in the previous section for the
description of liquid--vapour phase coexistence of a simple fluid
in a HS random porous medium. However, we first consider the bulk case, i.e., when $\rho_{0}^{*}=0$.
The fluid-fluid interaction $v_{11}(r)$ is taken in the form~(\ref{hol2.20}) with the attractive potential~$u_{11}(r)$~(\ref{hol2.39}).
In figure~\ref{Fig3}, one can see the liquid--vapour phase diagrams in coordinates $T-\rho_1$
for a bulk fluid obtained within different approximations.
The simulation results obtained in~\cite{Alvarez99} using the method of grand-canonical
Monte Carlo are shown for a comparison. As it is seen, the HTA approximation gives a good description only for low
temperature and leads to the overestimation at higher temperatures.
The BH approximation slightly improves the diagram, but still essentially overestimates the critical temperature.
As expected, the MSA approximation provides the best description of the phase diagram among the considered ones, although
the critical temperature is still higher than in the simulations.

\begin{figure}[!t]
\centerline{
\includegraphics [width=0.48\textwidth]  {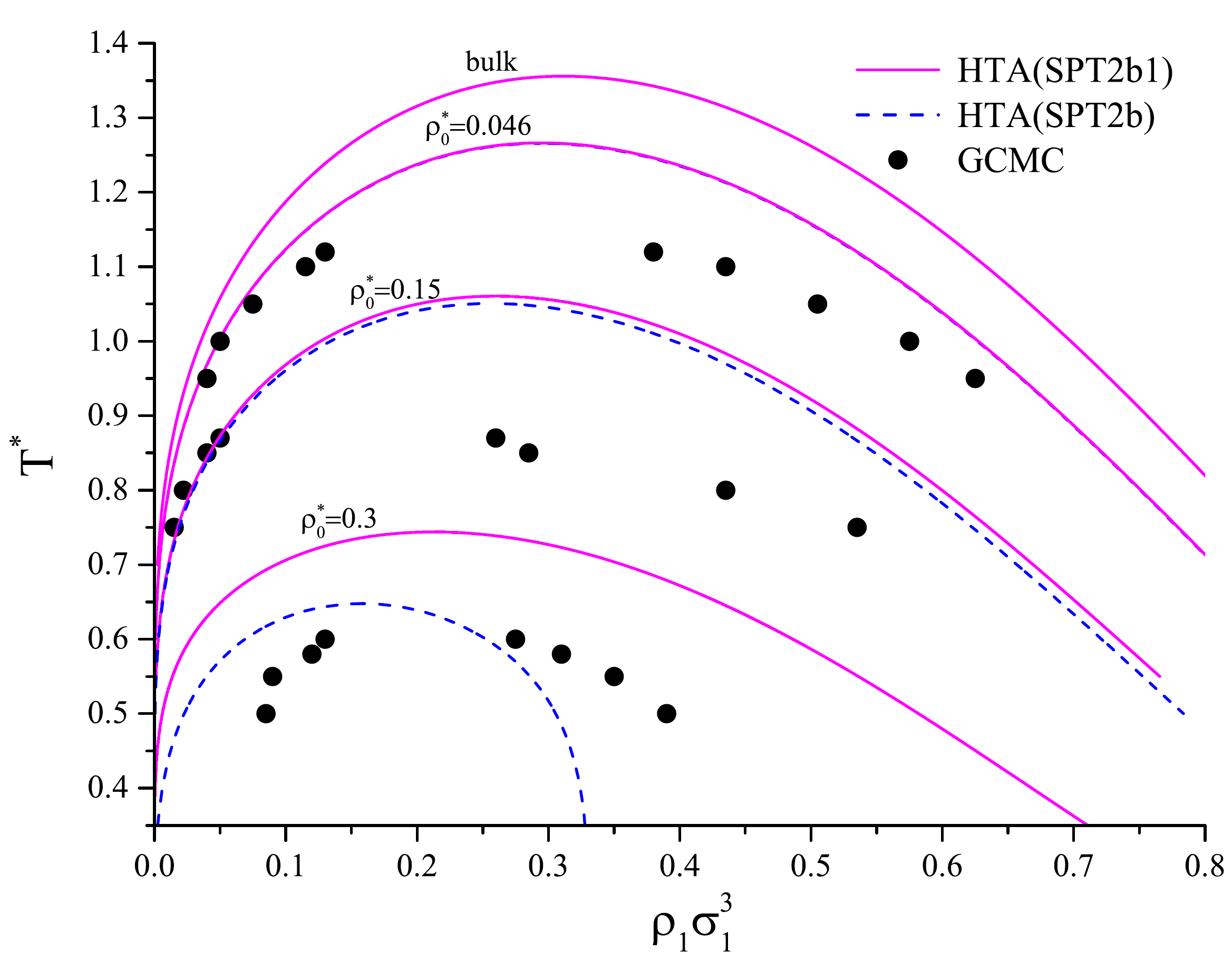}
\includegraphics [width=0.48\textwidth]  {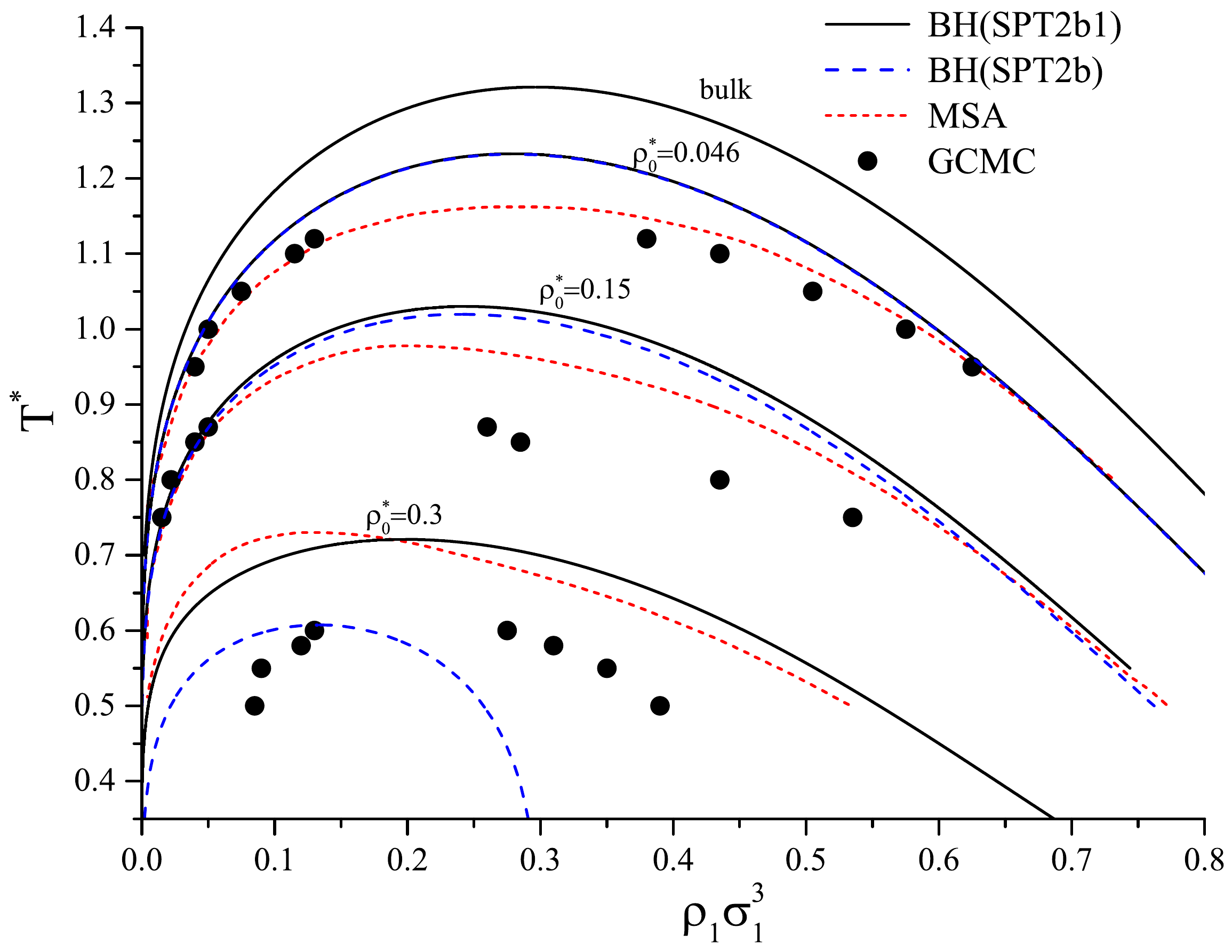}
}
\caption{(Color online) Liquid--vapour phase diagrams for the same fluid as in figure~\ref{Fig3},
but confined in a matrix of different densities $\rho_{0}^{*}=\rho_{0}\sigma_{1}^{3}=0.046$,
$0.15$ and $0.30$, where $T^{*}=k_\textrm{B}T/\epsilon$.
Left-hand panel: dashed lines correspond to the HTA approximation combined with SPT2b,
solid lines~--- HTA combined with SPT2b1, symbols~--- GCMC results taken from~\cite{Alvarez99}.
Right-hand panel: dashed lines correspond to the BH theory combined with SPT2b,
solid lines~--- BH theory combined with SPT2b1, dotted lines~--- MSA approximation,
symbols~--- the GCMC results taken from~\cite{Alvarez99}.}
\label{Fig4}
\end{figure}

\begin{figure}[!t]
\centerline{
\includegraphics [width=0.5\textwidth]  {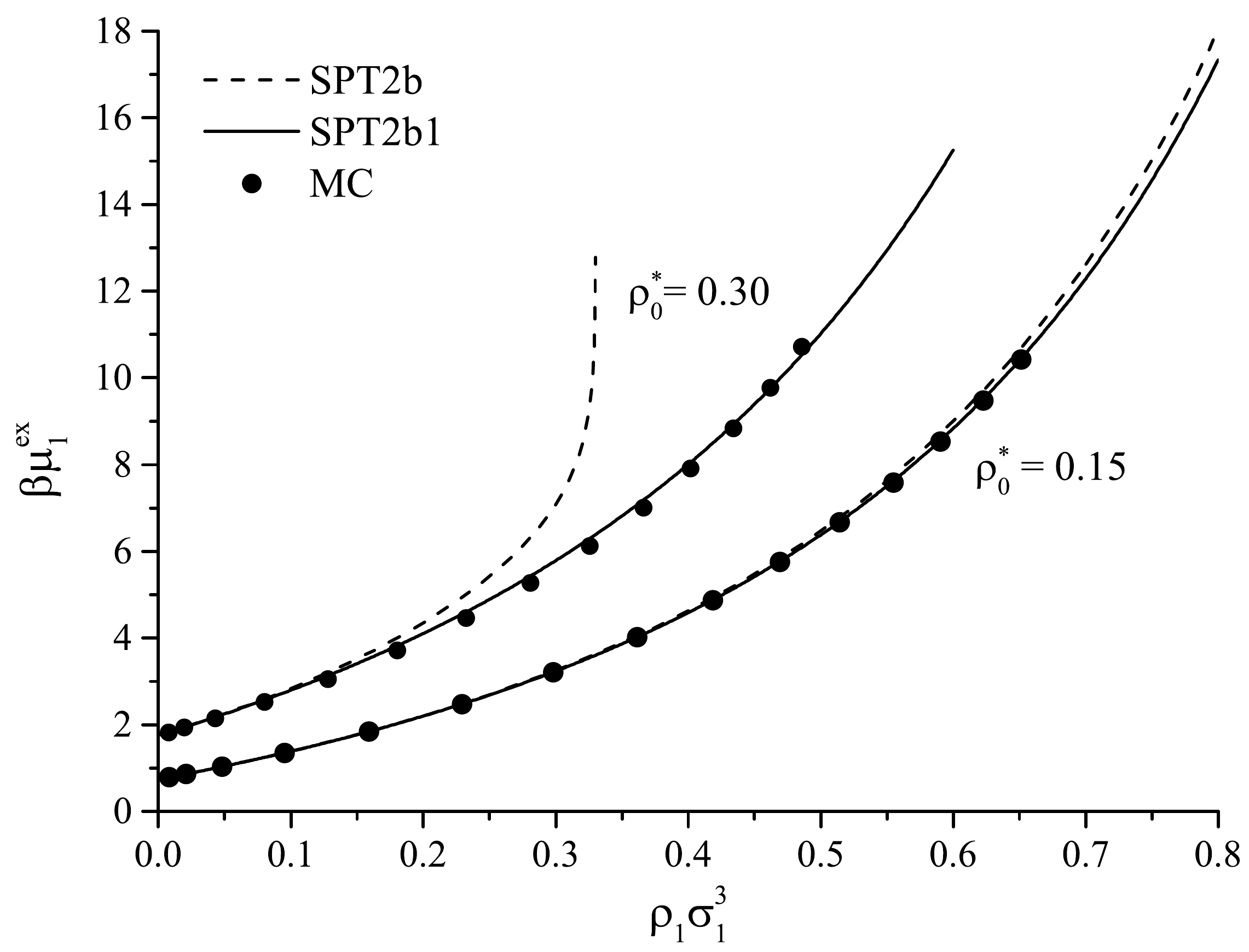}
}
\caption{The chemical potential for a HS fluid confined in a HS matrix of densities $\rho_{0}^{*}=\rho_0\sigma_1^3=0.15$ and $0.30$,
and the ratio $\tau=\sigma_1/\sigma_0=1.0$. A comparison between the results of SPT2b~(dashed lines), SPT2b1~(solid lines)
and the grand-canonical Monte Carlo simulations performed in the present study~(symbols).}
\label{MC_SPT}
\end{figure}

Considering a fluid in a matrix within the model applied in our study,
one should take into account that there is no attractive interaction between fluid and
matrix particles. Thus, the only effect of the matrix is to confine the fluid in the void volume
formed between matrix particles. Therefore, the most relevant parameters, which determine fluid properties,
are the matrix density $\rho^{*}_{0}$ or the corresponding matrix porosity $\phi_{0}$,
and the size ratio of fluid and matrix particles $\tau=\sigma_{1}/\sigma_{0}$.
In figure~\ref{Fig4}, we present the liquid--vapour phase diagrams obtained
for a fluid in matrices of different densities $\rho_{0}^{*}=0$ (bulk), $0.046$, $0.15$ and $0.30$.
In this case, the fluid and matrix particles are of equal sizes, i.e., $\tau=1$.
For this purpose, we use the theoretical approaches considered in the previous
section, i.e., the HTA and BH approximations in combination
with the reference system obtained within the SPT2b and SPT2b1 approaches.
 The computer simulation results taken from~\cite{Alvarez99} are
presented in figure~\ref{Fig4} for comparison. Similarly to the bulk case, one can see the MSA results for the considered
systems~\cite{Kierlik97}. It is observed that for a low matrix density $\rho_{0}^{*}=0.046$,
the coexistence curves calculated using the SPT2b and SPT2b1 approximations almost coincide.
With an increase of the matrix density up to $\rho_{0}^{*}=0.15$,
the difference between the results obtained with SPT2b and SPT2b1 becomes more distinguishable.
And finally, for the high matrix density $\rho_{0}^{*}=0.3$, the diagrams differ essentially
in these approximations. Moreover, the results obtained with the use of the SPT2b are rather anomalous,
since they are far from any other approximation.
This anomaly can be explained by the divergence contained in
the expressions for the chemical potential and pressure of a fluid when $\eta_{1}\rightarrow\phi$~\cite{Hol12,Hol13},
and which become important at high matrix densities. To illustrate this problem, the dependencies of the chemical potential
of a HS fluid in a HS matrix of densities $\rho_{0}^{*}=0.15$ and $\rho_{0}^{*}=0.3$, in comparison with the results of
grand-canonical Monte Carlo are shown in figure~\ref{MC_SPT}. It is clearly seen that for the matrix density $\rho_{0}^{*}=0.3$
the chemical potential of a fluid becomes wrong at the densities $\rho_{1}^{*}>0.2$ and tends to infinity around~$0.33$.
The latter value corresponds to the $\eta_{1}\approx\phi$, where the divergence is expected. However, this is not
the case for $\rho_{0}^{*}=0.15$, where the SPT2b approximation is close to the result of SPT2b1 approximation.
On the other hand, it is observed that the SPT2b1 approximation perfectly fits the simulation results for the both
matrix densities up to the highest values of fluid densities. Therefore, STP2b1 should be considered as the best choice
for the description of a HS fluid in HS matrix. Consequently, hereafter we restrict ourselves only to this approximation for the reference system.

All the approximations HTA, BH and MSA correctly reproduce the basic trends of
the behaviour of liquid--vapour coexistence curves of a fluid in a matrix,
i.e., a decrease of matrix porosity (or an increase of the matrix density) leads to a critical point shift
toward lower fluid densities and lower temperatures, simultaneously the phase diagram becomes narrower~(figure~\ref{Fig4}).
Furthermore, all of the considered approximations give only one critical point.
A comparison of the diagrams obtained using the approximations presented in our paper with computer simulations~(figure~\ref{Fig4}) shows that the
inclusion of the second term in the BH theory essentially improves the
description of coexistence curves. The computer simulations data are in
semiquantitative agreement with the theoretical prediction based on the MSA and BH
approximation. It is worth mentioning that in contrast to MSA, which
is mostly numerical approach, the BH approximation is more an analytical theory.

\begin{figure}[!b]
\centerline{
\includegraphics [width=0.5\textwidth]  {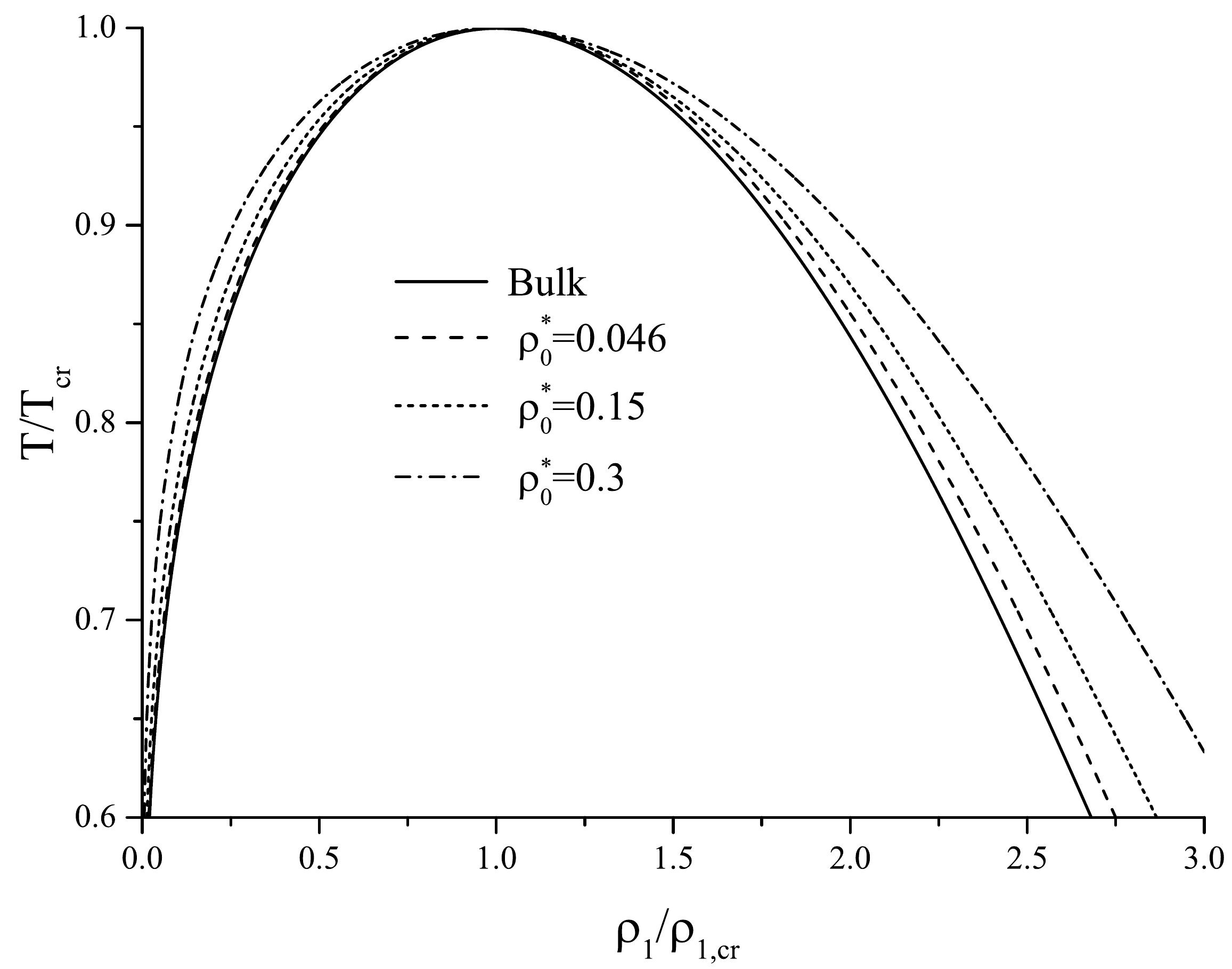}
}
\caption{Liquid--vapour coexistence diagrams in terms of reduced temperature $T/T_\textrm{cr}$ and reduced density $\rho_{1}/\rho_\textrm{1,cr}$ for
 different matrix densities $\rho_{0}^{*}$ and for the same fluid model as in figures~\ref{Fig3}--\ref{Fig4}. All curves are calculated within the framework of the BH theory with the description of the reference system within SPT2b1 approximation.}
\label{Fig5}
\end{figure}

The phase diagrams obtained for a fluid in the matrices of different densities are presented in the reduced units as a plot $T/T_\textrm{cr}$ versus
$\rho_{1}/\rho_\textrm{1,cr}$ in figure~\ref{Fig5}. In such a way, it makes possible to check whether the theory developed for
a fluid confined in a HS matrix satisfies the law of the corresponding states.
As one can see in the figure, the considered diagrams are rather close to each other, except
the case of high matrix density $\rho_{0}^{*}=0.30$. A general trend of the phase diagrams becomes broader
when the reduced temperature decreases.
\begin{figure}
\centerline{
\includegraphics [width=0.5\textwidth]  {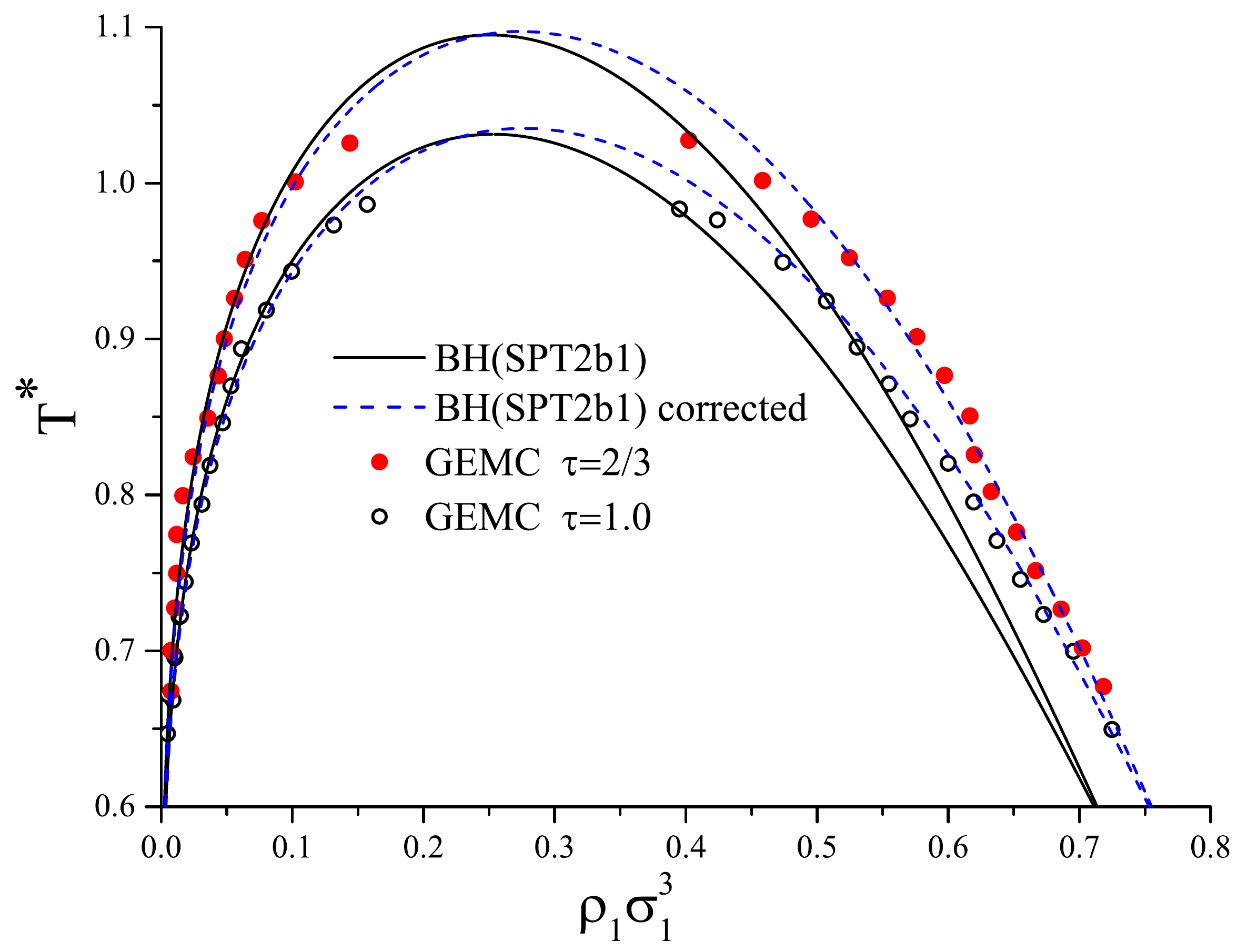}
}
\caption{(Color online) Liquid--vapour phase diagram for a Lennard-Jones fluid in a HS matrix for different size ratios $\tau=\sigma_{1}/\sigma_{0}$,
but at fixed porosity $\phi_{0}=0.95$. Theoretical predictions are given using the BH theory with the description of the reference system
within framework of the SPT2b1 approach. Solid lines~--- reference system with a hard core size $d_1(T)=\sigma_1$. The dashed lines~---
corrected results with $d_1(T)$ defined according to the Barker-Henderson formula~(\ref{hol2.43}).
Circles~--- GEMC results taken from \cite{Brennan02}.}
\label{Fig6}
\end{figure}

Now, we consider the confinement effect of a matrix on a fluid by varying the size ratio
of the fluid and matrix particles $\tau=\sigma_{1}/\sigma_{0}$ at a fixed porosity.
At the same time, we compare the theoretical results with computer simulations data obtained in~\cite{Brennan02}
using the method of Gibbs-ensemble Monte Carlo~(GEMC) for the conventional Lennard-Jones~(LJ) potential
\begin{equation}
\label{hol2.40}
{\it v}_{11}(r)=4\epsilon\left[\left(\frac{\sigma_{1}}{r}\right)^{12}-\left(\frac{\sigma_{1}}{r}\right)^{6}\right],
\end{equation}
truncated at the distance $r_\textrm{c}=2.5\sigma_{1}$. Since in our study the reference system is taken as a HS system,
the repulsive part of LJ potential is a hard core potential. Thus, according to the form (\ref{hol2.20}), we use
\begin{eqnarray}
{\it v}_{11}(r)=\left\{\begin{array}{ll}
\infty,&r<\sigma_{1}\,,\\
4\epsilon\left[\left(\sigma_{1}/r\right)^{12}-\left(\sigma_{1}/r\right)^{6}\right],& \sigma_{1}<r<2.5 \sigma_{1}\,,\\
0,& r>2.5 \sigma_{1}\,.
\end{array}\right.
\label{hol2.41}
\end{eqnarray}

In figure~\ref{Fig6} (solid lines) there is presented a comparison of theoretical results calculated for the fluid-fluid potential~(\ref{hol2.41}) with computer simulations~\cite{Brennan02}~(symbols) for the LJ potential~(\ref{hol2.40}) at $\tau=1$ and $\tau=3/2$, but for
the fixed matrix porosity $\phi_{0}=0.95$ (or the matrix packing fraction $\eta_{0}=0.05$).
As one can see, the form of the phase diagrams obtained from the BH theory is rather close to the data of simulations,
although for the both cases of $\tau$ they are notably narrower. This deviation is systematic and anticipated, since the simulations are
performed for the conventional LJ potential, which has a soft core, while we use the reference system as a hard core fluid.
According to (\ref{hol2.40}), the repulsive potential is as follows:
\begin{eqnarray}
\varphi_{11}(r)=\left\{\begin{array}{ll}
4\epsilon\left[\left(\sigma_{1}/r\right)^{12}-\left(\sigma_{1}/r\right)^{6}\right] ,& r<\sigma_{1}\,, \\
0,& r>\sigma_{1}\,.
\end{array}\right.
\label{hol2.41'}
\end{eqnarray}
It is acceptable to substitute this repulsive part of the LJ potential by a hard core, but instead of the diameter of HS particles $\sigma_1$,
one should take an effective diameter $d_1$ which is somewhat smaller than $\sigma_1$ and depends on the fluid temperature.
To take into account a correct value of the effective diameter $d_1$ we use one of the successful and the most popular relations
proposed by Barker and Henderson for a LJ fluid~\cite{BarkerHen67}:
\begin{equation}
\label{hol2.43}
d_{1}^\textrm{BH}(T)=\int_{0}^{\sigma_{1}}\left\{1-\exp\left[-\beta\varphi_{11}(r)\right]\right\}{\rm d}r.
\end{equation}
Simple calculations of $d_1^\textrm{BH}(T)$ depending on temperature show that it does not vary too heavily.
For instance, for the temperature $T^{*}=1.1$,  the effective diameter is $d_{1}^\textrm{BH}=0.9711$
and for $T^{*}=0.6$, the diameter is $d_{1}^\textrm{BH}=0.9815$.
However, it still can have a strong effect on the thermodynamics of the system, hence on the curves of the liquid--vapour coexistence.
Therefore, we need to substitute the diameter of HS particles $\sigma_1$ by $d_{1}^\textrm{BH}(T)$ in every place where it is needed in the expressions used for the reference system, i.e., the terms containing $\sigma_1$ or depending on it should be modified.
The corrected expressions for the chemical potential and the pressure for a fluid in a HS matrix
are used and calculated depending on the temperature. Again using the BH theory in combination with the SPT2b1 approximation,
the phase diagrams of liquid--vapour transition are obtained. The results of this correction is presented as
dashed lines in figure~\ref{Fig6}, and, as one can see, the theoretical curves coincide very well with the computer simulation data.

\begin{figure}[!t]
\centerline{
\includegraphics [width=0.48\textwidth]  {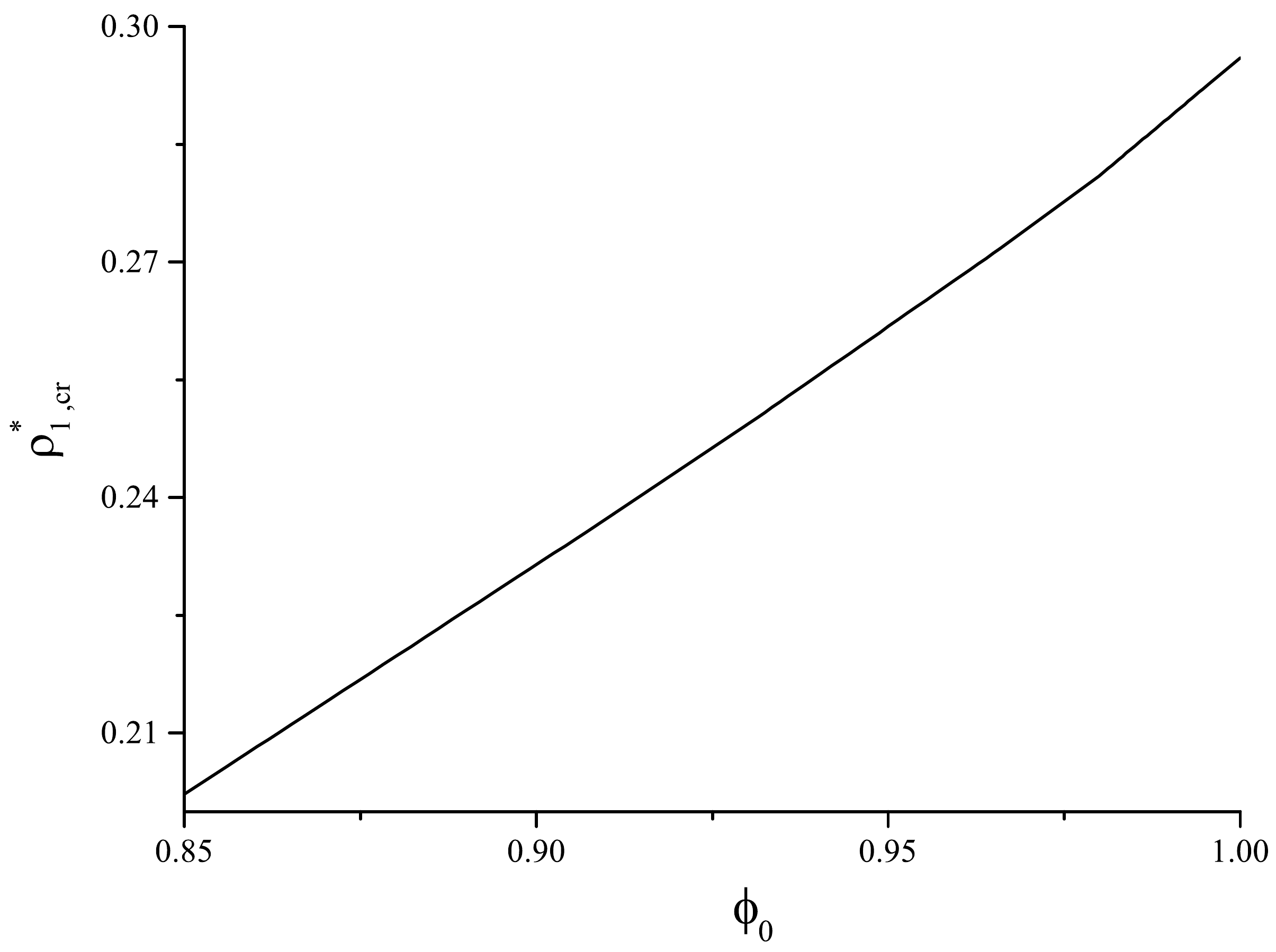}
\includegraphics [width=0.48\textwidth]  {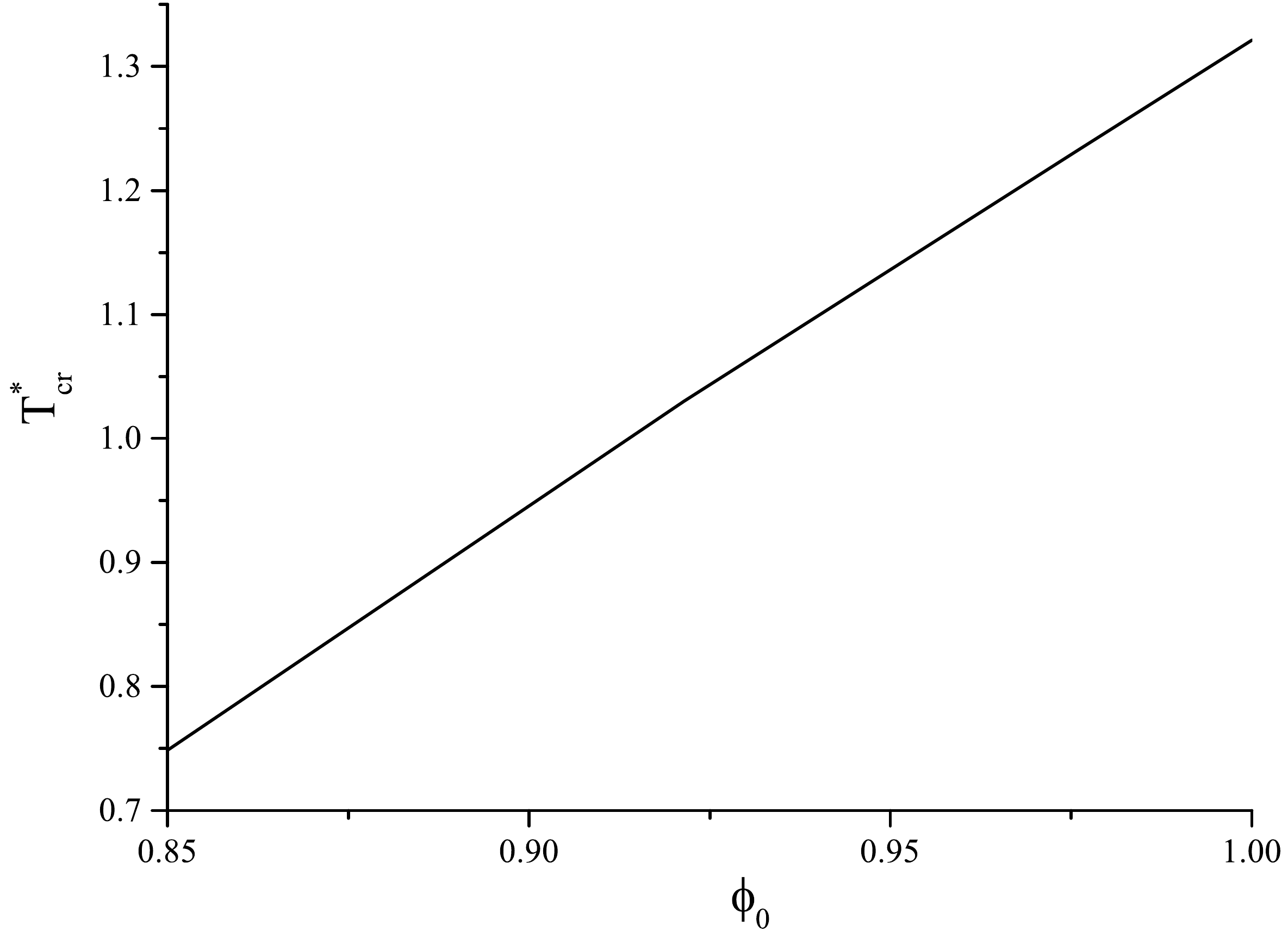}
}
\caption{The dependencies of the critical density $\rho_\textrm{1,cr}^{*}$ and critical temperature $T_\textrm{cr}^{*}$ on the matrix porosity $\phi_{0}$ for the same model as in figures~\ref{Fig3}--\ref{Fig4}.}
\label{Fig7}
\end{figure}

It is worth noting that there is a quicker and more efficient way to improve the present
phase diagrams. Our preliminary calculations show that coexistence curves change negligibly along the $T$-axis, and
 most deviations take place along the $\rho_1$-axis. This is mainly related to the effect of an excluded volume, which
depends on $\eta_1=\pi d_{1}^{3}\rho_1/6$ and is overestimated in the
case of the reference system with a hard sphere size $d_1=\sigma_1$.
Therefore, to improve our results for the conventional LJ fluid, the packing fraction should be replaced by $\eta_1^\textrm{BH}=\pi [d_{1}^\textrm{BH}(T)]^3\rho_1/6$,
and this is equivalent to the rescaling of the fluid density as:
\begin{equation}
\label{hol2.44}
\rho_1^\textrm{BH}=\rho_{1}\left[\frac{\sigma_1}{d_{1}^\textrm{BH}(T)}\right]^{3}.
\end{equation}
The correction made in such a way allows us to obtain the diagrams which are practically
equivalent to those shown in figure~\ref{Fig6}.

Finally, in figure~\ref{Fig7} we present the critical temperature $T_\textrm{cr}^{*}$
and the critical density $\rho_\textrm{1,cr}^{*}$ as a function of porosity $\phi_{0}$ for the
fluid confined in a HS matrix within the framework of the model considered in figure~\ref{Fig4}.
One can see here the effects similar to those observed in figure~\ref{Fig4} and
figure~\ref{Fig6}, i.e., with a decrease of porosity $\phi_{0}$, the critical
point shifts toward lower temperatures and densities. In figure~\ref{Fig7} we
do not present the results for the model discussed in figure~\ref{Fig6}, since
the results in figure~\ref{Fig4} and figure~\ref{Fig6} are shown for a
different model of the fluid. However, we should remark that the critical
temperature decreases with an increase of $\tau$ at the fixed porosity, and the change
of the critical density is very small. For example, from the results presented in
figure~\ref{Fig6} it is found that for $\phi_{0}=0.95$ and $\tau=2/3$, the critical temperature $T^{*}_\textrm{cr}=1.095$ and
the critical density is $\rho_\textrm{1,cr}^{*}=0.273$, while for $\tau=1$, the critical temperature is $T_\textrm{cr}^{*}=1.031$ and
the critical density is $\rho_\textrm{1,cr}^{*}=0.275$. If we estimate this in the limit $\tau\rightarrow0$,
one can see that the critical temperature $T^{*}_\textrm{cr}$ shifts to $T_\textrm{cr}^{*(\text{bulk})}=1.209$ and
the critical density to $\rho_{1cr}^{*}\rightarrow \rho_{1cr}^{*(\text{bulk})}$
$\phi_{0}=0.305\times0.95=0.29$. $\rho_{1cr}^{*(\text{bulk})}$ and $T_\textrm{cr}^{*(\text{bulk})}$ correspond to the values of the critical parameters of a bulk fluid.
\section{Conclusions}

In this paper, the Barker-Henderson~(BH) perturbation theory is generalized for the
a Lennard-Jones fluid confined in a random porous matrix. As the reference system, a hard sphere
fluid in a hard sphere matrix is chosen. To describe the reference system, the extension of
the scaled particle theory~(SPT) is used, and two corresponding approximations are tested.
It is shown that the SPT2b1 approximation, which was developed recently, makes it possible
to achieve a very accurate description of the thermodynamics of confined hard sphere systems.
Combining the SPT approach with the BH theory, the expressions for the chemical potential
and the pressure of a simple fluid in a hard sphere matrix are derived. Based on the obtained
expressions, the phase diagrams of liquid--vapour transition are calculated and compared with other
theoretical approaches such as the high-temperature approximation and the mean-spherical approximation.
A comparison of our results with computer simulation data found in the literature is made as well.
Different matrix porosities as well as the size ratios of fluid and matrix particles are considered in the present paper.
The proposed extension of the BH theory for the case of a fluid in a matrix provides a good qualitative agreement with the
computer simulations, and in some situations it provides a marvelous quantitative agreement, as it is observed in the case of low matrix porosities
and at temperatures which are not very close to the critical point.
The theory correctly reproduces the basic effects of porous media on
the liquid--vapour phase coexistence of simple fluids, i.e., with a decrease of
porosity, the critical point shifts toward lower fluid densities and lower
temperatures. It is also observed that for a fixed matrix porosity, but
for variable sizes of matrix particles, the critical temperature increases if the
size of matrix particles becomes larger and moves to the value of critical
temperature of a bulk fluid, while the critical density
changes weakly, and in the limit $\tau\rightarrow0$, it moves to the bulk
critical value normalized by the porosity $\phi_{0}$.

The approach developed in this paper can be extended to the case of more complex fluid systems
in a confinement. In future, we plan to generalize the BH theory for anisotropic fluids in
random porous media. Our very recent investigation~\cite{HolShmotPat} connected
with the extension of the Van der Waals theory to the case of anisotropic fluids in
random porous matrices shows that due to anisotropic interactions, the orientational
order causes a competition between isotropic and anisotropic interactions,
and the effect of a matrix can essentially modify the liquid--vapour phase diagram.

\begin{figure}[!t]
\centerline{
\includegraphics [width=0.49\textwidth]  {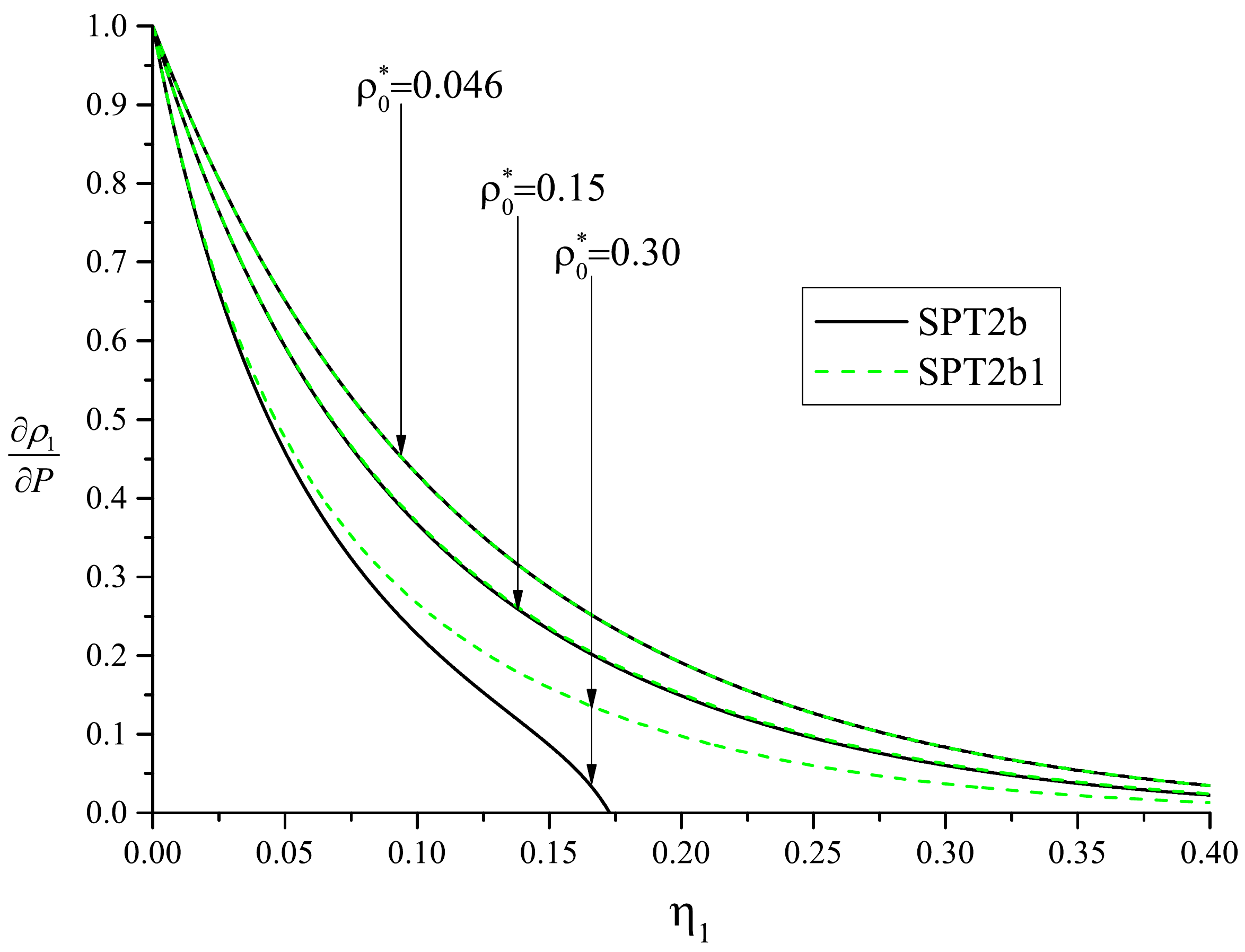}
\includegraphics [width=0.49\textwidth]  {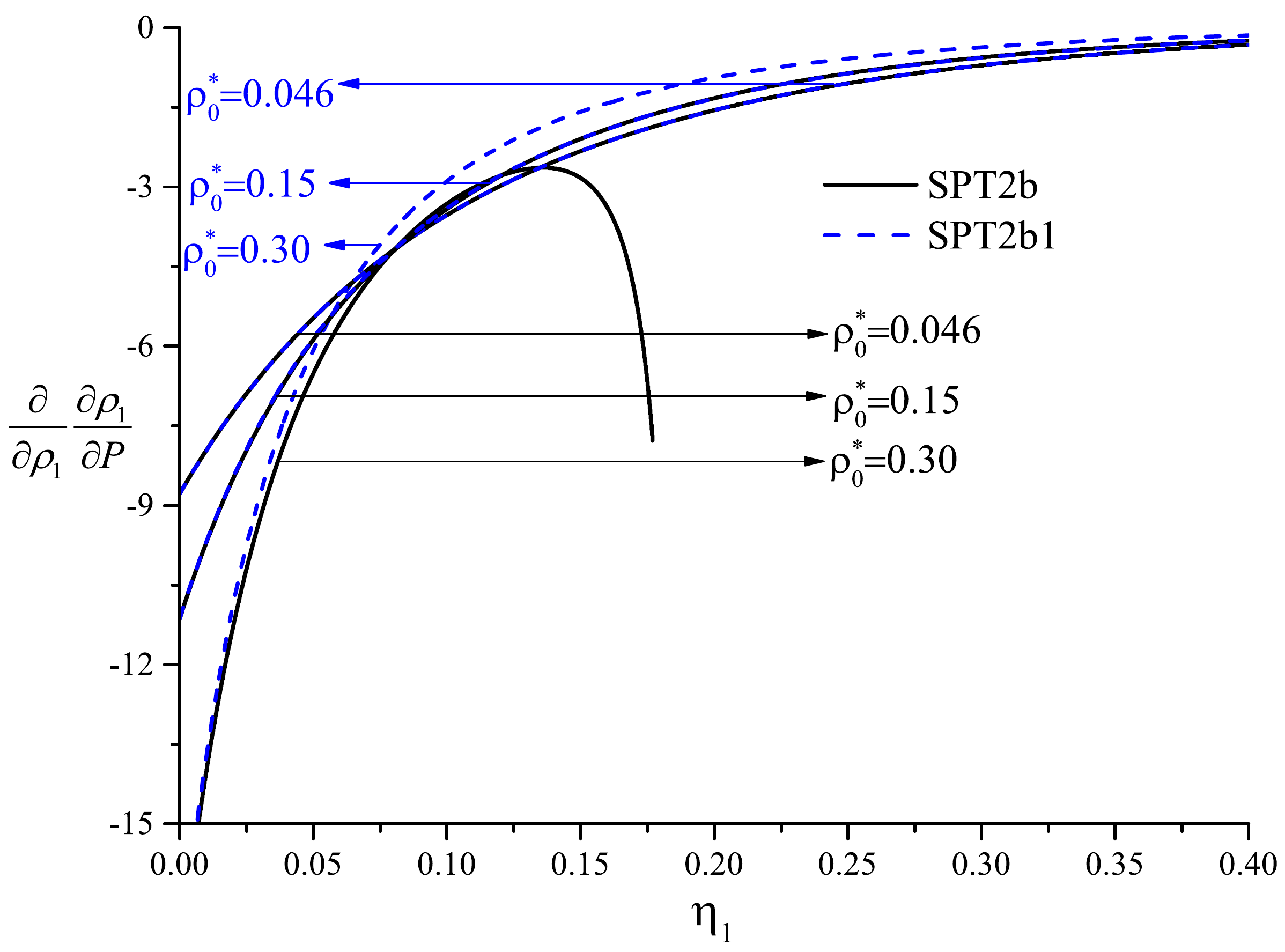}
}
\caption{(Color online) Isothermal compressibility $\left(\frac{\partial\rho_{1}}{\partial P}\right)_{T}$ and the derivative
$\frac{\partial}{\partial\rho_{1}}\left(\frac{\partial\rho_{1}}{\partial P}\right)_{T}$ for a HS fluid in a HS matrix calculated within SPT2b and SPT2b1
approximations.}
\label{Fig8}
\end{figure}

\appendix

\section{Isothermal compressibility for a HS fluid in HS
matrix}

Here, we present the expression for the isothermal compressibility
$\left(\frac{\partial\rho_{1}}{\partial P}\right)_{T}$ and the derivative \linebreak
$\frac{\partial}{\partial\rho_{1}}\left(\frac{\partial\rho_{1}}{\partial
P}\right)_{T}$ for a HS fluid in a HS matrix obtained within the SPT2b and SPT2b1
approximations. Using simple differentiations of the expressions
(\ref{hol2.16}) and (\ref{hol2.19}), one obtains
\begin{equation}
       \left(\frac{\partial \rho_{1}}{d P}\right)^\textrm{SPT2b}  =  {\left[ -\frac{ {\phi_{0}}/{\phi }-1
 }{1-\eta_{1}/\phi_{0}} +\frac{\phi_{0}}{\phi}
 \frac{1}{\left(1-\eta_{1}/\phi_{0}\right)^{2}}  +A\frac{
 \eta_{1}/\phi_{0}}{\left(1-\eta_{1}/\phi_{0}\right)^{3}}  +2B\frac{
 (\eta_{1}/\phi_{0})^{2}}{\left(1-\eta_{1}/\phi_{0}\right)^{4}}  \right]}^{-1},
 \label{hol A.1}
 \end{equation}
 \begin{align}
      \frac{\partial}{\partial\rho_{1}}\left(\frac{\partial\rho_{1}}{\partial P}\right)_{T}^\textrm{SPT2b}  &=- \left[ -\frac{
 {\phi_{0}}/{\phi }-1  }{ \phi_{0} (1-\eta_{1}/\phi_{0})^{2}}
 +\frac{2}{\phi(1-\eta_{1}/\phi_{0})^{3}} +  A\frac{1+2\eta_{1}/\phi_{0}}{ \phi_{0}(1-\eta_{1}/\phi_{0})^{4}}
 + 4B\frac{\eta_{1}/\phi_{0}+(\eta_{1}/\phi_{0})^{2}}{\phi_{0}
 (1-\eta_{1}/\phi_{0})^{5}}  \right]
 \nonumber\\ &
 \times  \left[ -\frac{
 \frac{\phi_{0}}{\phi }-1  }{1-\eta_{1}/\phi_{0}} +\frac{\phi_{0}}{\phi}
 \frac{1}{(1-\eta_{1}/\phi_{0})^{2}}
 +A\frac{
 \eta_{1}/\phi_{0}}{(1-\eta_{1}/\phi_{0})^{3}}+2B\frac{(\eta_{1}/\phi_{0})^{2}}{(1-\eta_{1}/\phi_{0})^{4}}  \right]^{-2},
 \label{hol A.2}
 \end{align}
 \begin{equation}
       \left(\frac{\partial\rho_{1}}{\partial P}\right)_{T}^\textrm{SPT2b1}  =  {\left[ \frac{ 1  }{1-\eta_{1}/\phi} +
 \frac{\eta_{1}/\phi_{0}}{\left(1-\eta_{1}/\phi_{0}\right)^{2}}  +A\frac{
 \eta_{1}/\phi_{0}}{\left(1-\eta_{1}/\phi_{0}\right)^{3}}  +2B\frac{
 (\eta_{1}/\phi_{0})^{2}}{\left(1-\eta_{1}/\phi_{0}\right)^{4}}  \right]}^{-1},
  \label{hol A.3}
 \end{equation}
 \begin{align}
 \frac{\partial}{\partial\rho_{1}}\left(\frac{\partial\rho_{1}}{\partial P}\right)_{T}^\textrm{SPT2b1}  &=-   \left[ -\frac{ 1  }{ \phi
(1-\eta_{1}/\phi_{0})^{2}}
 +\frac{1+\eta_{1}/\phi_{0}}{\phi_{0}(1-\eta_{1}/\phi_{0})^{3}} +
 A\frac{ 1+2\eta_{1}/\phi_{0}}{ \phi_{0}(1-\eta_{1}/\phi_{0})^{4}}
 +4B\frac{ \eta_{1}/\phi_{0}+(\eta_{1}/\phi_{0})^{2}}{\phi_{0}
(1-\eta_{1}/\phi_{0})^{5}}  \right]
\nonumber\\  &
\times
  \left[ \frac{
 1  }{1-\eta_{1}/\phi} +
 \frac{\eta_{1}/\phi_{0}}{(1-\eta_{1}/\phi_{0})^{2}}
 +A\frac{
 \eta_{1}/\phi_{0}}{(1-\eta_{1}/\phi_{0})^{3}}  +2B\frac{
 (\eta_{1}/\phi_{0})^{2}}{(1-\eta_{1}/\phi_{0})^{4}}  \right]^{-2},
  \label{hol A.4}
 \end{align}
where $A$ and $B$ are given in equation (\ref{hol2.10}).

The dependence of $\left(\frac{\partial\rho_{1}}{\partial P}\right)_{T}$ and
$\frac{\partial}{\partial\rho_{1}}\left(\frac{\partial\rho_{1}}{\partial
P}\right)_{T}$ on the fluid packing fraction
$\eta_1$ calculated within the SPT2b and SPT2b1 approximations for different matrix densities $\rho_{0}^{*}$ is shown
in figure~\ref{Fig8}. One can see that for low densities of $\rho_{0}^{*}$, the
results in the both approximations are nearly identical, but for $\rho_{0}^{*}=0.30$,
the results within SPT2b show an odd behaviour for $\eta_{1}$ larger than 0.15. 

\ukrainianpart

\title{Що таке рідина в невпорядкованому пористому середовищі: теорія збурень Баркера-Гендерсона}
\author{М.Ф. Головко, Т.М. Пацаган, В.І. Шмотолоха}
\address{
Інститут фізики конденсованих систем НАН України, вул. І.~Свєнціцького, 1, 79011 Львів, Україна
}

\makeukrtitle

\begin{abstract}
\tolerance=3000%
Застосовано теорію збурень Баркера-Гендерсона (БГ) для вивчення плину Ленарда-Джонса в невпорядкованій пористій матриці, сформованій твердими сферичними частинками. З метою опису системи відліку, яка необхідна для теорії збурень, було використано розвинення теорії масштабної частинки (ТМЧ). Останні досягнення у розвитку ТМЧ для твердокулькового плину в твердокульковій матриці дозволяють отримувати термодинамічні властивості в такій системі із високою точністю. Таким чином, нами поєднано теорію БГ з теорією ТМЧ та виведено вирази для хімічного потенціалу і тиску плину в матриці. Використовуючи отримані вирази та умови фазової рівноваги, побудовано фазові діаграми газ--рідина плину Ленарда-Джонса в твердокульковій матриці. Досліджено ефект пористості матриці і розміру матричних частинок. Показано, що зменшення пористості матриці понижує значення критичної температури і критичної густини плину, разом з тим, фазова діаграма звужується. Також спостережено, що збільшення розміру матричних частинок призводить до росту критичної температури. Зауважено, що результати теорії узгоджуються із даними комп'ютерного моделювання. Запропонований  теоретичний підхід може бути розвинутий до опису анізотропних рідин у твердокульковій матриці.
\keywords плини в невпорядкованих пористих середовищах, теорія збурення Баркера-Гендерсона, фазовий перехід газ-рідина, теорія масштабної частинки

\end{abstract}

\end{document}